\pdfoutput=1

\documentclass[a4paper,11pt]{article}  

\usepackage{a4wide}
\usepackage[utf8]{inputenc}
\usepackage[T1]{fontenc}
\usepackage[english]{babel}
\usepackage[babel]{csquotes}
\usepackage[english]{varioref}
\usepackage{amsmath,amsthm,mathrsfs}
\usepackage{amsfonts,amssymb}
\usepackage{listings}
\usepackage{braket,array,booktabs}
\usepackage[font=small,format=hang,labelfont={sf,bf}]{caption}
\usepackage{mathtools,empheq}
\usepackage{siunitx}
\usepackage[a4paper,left=2.5cm,right=2.5cm,top=3.5cm,bottom=3.5cm,
 bindingoffset=0mm]{geometry}
\usepackage[pdftex]{hyperref}
\usepackage[pdftex]{graphicx}
\usepackage{subcaption}

\hypersetup{
  pdfauthor = {Francesco Ricci, Giovanni Frosali},
  pdftitle = {A Symbolic Method for Nonlinear Two-Mass-Skate Analysis},
  pdfkeywords = {Bicycle dynamics, nonholonomic constraints, nonlinear dynamics, stability analysis},
  colorlinks = true,
  linkcolor = black,
  citecolor = black
}

\newcommand{\email}[1]{\href{mailto:#1}{\texttt{#1}}}

\newcommand{\specialcell}[2][l]{%
  \begin{tabular}[#1]{@{}l@{}}#2\end{tabular}}

\newenvironment{system}%
{\left\lbrace\begin{array}{@{}l@{}}}%
{\end{array}\right.}

\newcommand{\R}{\mathbb{R}}

\newcommand{\f}{\varphi}

\newcommand{\dpe}{\partial}

\let\ps\relax
\newcommand {\ps} {\,{\bf \cdot}\,}

\newcommand{\mc}[1]{\mathcal{#1}}
\newcommand{\mr}[1]{\mathrm{#1}}

\newcommand{\inca}[1]{\mathopen{[}{#1}\mathclose{)}}

\newcommand{\ina}[1]{\mathopen{(}{#1}\mathclose{)}}
\newcommand{\ol}[1]{\overline{#1}}
\newcommand{\wt}[1]{\widetilde{#1}}

\newcommand{\xp}{\dot x}
\newcommand{\yp}{\dot y}

\newcommand{\zo}{\omega}
\newcommand{\za}{\alpha}

\newcommand{\zt}{\theta}

\newcommand{\thetati}{\widetilde{\theta}}
\newcommand{\ztp}{\dot \theta}
\newcommand{\zap}{\dot \alpha}

\newcommand{\TT}{K}

\numberwithin{equation}{section}

\DeclarePairedDelimiter{\scal}{\langle}{\rangle}

\DeclarePairedDelimiter{\abs}{\lvert}{\rvert}

\theoremstyle{definition}

\theoremstyle{plain}

\newtheorem{prb}{Question}

\theoremstyle{remark}
\newtheorem{rmk}{Remark}[section]

\allowdisplaybreaks		

\author{
\textbf{Francesco Ricci}\\
Department of Mathematics\\
Imperial College London - Exhibition Road \\
London  SW7 2AZ \\
\email{f.ricci12@imperial.ac.uk} \\
\textbf{Giovanni Frosali} \\
Dipartimento di Matematica e Informatica ``U. Dini''\\
Universit\`a di Firenze - Via S.Marta, 3 \\
I-50139 Firenze, Italy \\
\email{giovanni.frosali@unifi.it}
}

\begin{document}

\title{A Symbolic Method for the Analysis of a Nonlinear Two-Mass-Skate Model}

%

%
%

\maketitle

\begin{abstract}
Multibody systems usually give rise to complex nonlinear dynamics, and the bicycle is not an exception. Even a simple model as the Two-Mass-Skate presents a long expression of the kinetic energy, making difficult to write explicitly the equations of motion. Instead of linearising or approximating the model, we will overcome this issue by using a functional expression of the kinetic energy. With introduction of appropriate nonlinear functions, the equations of motion are written in a form that can be easily handled despite their complexity. A stability analysis of the dynamics is then conducted.
\end{abstract}

\section{Introduction}

The bicycle is an example of multibody system which is widely diffused and exhibits a large number of interesting behaviours. The stability of a bicycle has interested many researchers for more than two centuries. The tendency of the system to reach an asymptotic equilibrium when perturbed makes the bicycle easy to ride, and many papers has been related to show the existence of such stable position. A detailed review of the works dedicated to this argument can be found in~\cite{meij:history}, where the most relevant papers are presented with useful comments.

In particular, one pioneering article is the 1899 paper by Whipple~\cite{whipple:bicycle}.  He derived a set of nonlinear equations for a bicycle model in order to study its stability. These equations of motion has been used in many papers by different authors who studied the dynamics of a bicycle. However, in order to analyse the stability properties of the system, the equations were usually linearised about the vertical position characterising the forward rectilinear motion.

For instance, in order to overcome the complexity of the equations, Whipple presented a linearisation about the rectilinear motion assuming that the rider does not use the handles. This case emphasises the self-stability of the bicycle, that is, the property of maintaining a stable position without the aid of the driver.

The self-stability of a bicycle has recently been studied more thoroughly, as the rideability of the system controlled by a rider is deeply connected to the self-stability exhibited by a riderless bicycle. For example, in~\cite{klein:bicycle} and~\cite{jones:bicycle} it is shown that particular experimental bicycles which do not present self-stability are difficult for a person to ride. The causes behind this phenomenon are not completely evident, but if it has been widely believed that gyroscopic and trail effects play an important role in bicycle stability, where the trail is given by the distance between the front ground contact point and the instersection of the steering axis with the ground. For instance, this idea is supported by several authors, as  in~\cite{cossalter:moto} and~\cite{cossalter:moto2}, who analysed directly the forces acting on the system and showed how gyroscopic effects and trail enhance its stability. However, this opinion has been completely changed in~\cite{meijaard:science}, where it is shown that a riderless two-wheeled vehicle  can be self-stable even without trail or gyroscopic effects. In support of this hypothesis, the authors also realised and studied a model, called Two-Mass-Skate (TMS).

The starting point for developing this system was the $25$ parameters model introduced in~\cite{whipple:bicycle}, which they reduced to a theoretical TMS characterised by only $8$ parameters. Consequently, the equations of motion for this system have a simpler expression as many terms are set to zero. Considering this simple system of equations, they were able to analyse the self-stability of the model. In particular, by taking a constant forward speed and assuming small lean and steering angles, they obtained a pair of coupled second-order differential equations for these two angles, and showed that, after small perturbations, the motion of the TMS exhibits asymptotically stability. To remark this result, they also realised a physical system on the basis of the assumptions taken for the theoretical one, and directly showed how the numerical simulations predict the behaviour of the real system.

On the other hand, in the paper it is also highlighted that particular technical choices had been taken in the realisation of the experimental TMS. For instance, they pointed out that the realisation of the ground contact represented a major concern, and that a specific aluminium wheel was required in order to have good stability properties. Therefore, it seems that the self-stability of the bicycle is independent of gyroscopic and caster effects from a theoretical point of view, that is, when there is no friction in the system and the energy is conserved. However, in a dissipative physical setting these forces may become necessary to the self-stability of the real system.

Despite these arguments, the TMS remains an interesting toy model for which it is possible to obtain a set of equations of motion that can be quite easily used to get results about the stability of the system. However, we believe that the linearised theoretical TMS presented in~\cite{meijaard:science} is probably an excessive idealisation of its physical counterpart, hence in this paper we first develop a model that combines the simple kinematic structure of the TMS with the geometric effects due to the distributed masses typical of a real system. Then, we derive the equations of motion taking into account the nonlinearity characteristic of the model, without linearising the equations of motion. Clearly, the equations found in this way present a more complex and richer structure with respect to their linearised versions, but we set up an interesting way to deal with the long expressions and get a deeper insight about the system. 
 
Therefore, the aim of this article is to obtain a set of nonlinear equations that accurately describe the TMS dynamic. In particular, we will impose no restrictions on the geometry of the rear and front frames, considering also their distributed masses, while we will still assume that both the wheels are material points without moments of inertia, and that the trail is set to zero. We will show that the behaviour of this new theoretical system generally agrees with the one of the original TMS, although, as expected, there are some significant differences when the initial configuration of the system is not close to the one used in the linearisation.

The paper is organised as follows. We first define the model from a geometrical point of view, characterising all the parameters which identify the system. Then we consider its kinematics and we introduce the nonholonomic constraints which determine the motion of the TMS on the ground. We will also refer to the general form of these constraints for a bicycle with toroidal wheels, and show how the simpler constraints of the TMS can be recovered from their full expression. To obtain the Lagrangian of the system, in Section~\ref{sec:lagrangian} we first write the kinetic energy of our TMS but, due to its complexity, we introduce a symbolic approach to deal with it. Using this symbolic approach, in Section~\ref{sec:motion} we derived the equations of motion for the TMS and then we obtain two special solutions for the system, namely, the forward motion and the circular motion. Finally, we report numerical simulations obtained from the nonlinear equations derived in the paper, comparing the behaviours of our system with those of the original TMS.

\section{Geometry and kinematics of the model}

\begin{figure}[htb]
\begin{center}
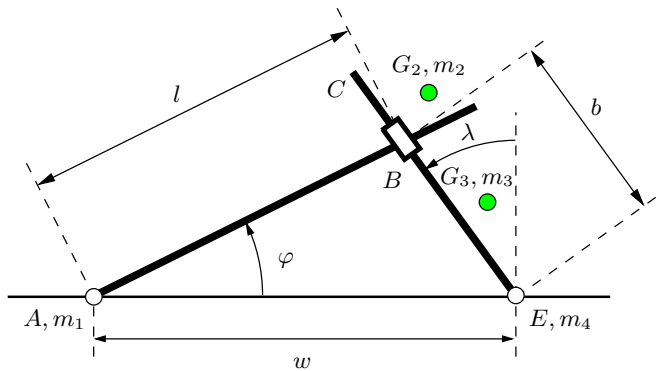
\caption{Two-Mass-Skate bicycle in the trivial configuration. The rear and front frames are linked together with a hinge which leaves the front frame free to rotate around the steering axis. The wheels have point masses, while the rear and front frames have distributed masses.}
\label{fig:TMS_model}
\end{center}
\end{figure}

The main feature of a Two-Mass-Skate bicycle is that the total mass of the bike is reduced to two point masses, one attached to the rear frame and one attached to the front frame, while the nonholonomic ground contacts are provided by two skates, as described in~\cite{meijaard:science}. Here we assume that the wheels have point masses and that both frames have distributed masses as well. We still call our model Two-Mass-Skate bicycle referring to the two distributed masses. Figure~\ref{fig:TMS_model} shows the TMS in the \emph{trivial configuration}, that is, the system is completely contained in a vertical plane normal to the ground. In the following we refer to the terminology introduced in~\cite{caim_simai2012}.

With labels as in Figure~\ref{fig:TMS_model}, let~$m_1$ and~$m_4$ be the masses of the rear and front wheels, respectively, $m_2$ the mass of the rod $AB$, and $m_3$ the mass of the rod $CE$, also called steering handle. Because the wheels have no physical extension, they both coincide with one endpoint of the corresponding rod, namely $A$ and $E$. Hence, we use the same labels even if they are different parts of the model. We remark that, while the wheels have no moments of inertia, both the frames have associated moment of inertia tensors. Having considered distributed masses for the frames, this model gives a closer description of the experimental TMS defined in~\cite{meijaard:science}.

Then, we define the \emph{wheelbase}~$w$ as the distance between the two contact points, the \emph{trivial pitch angle}~$\f$ as the angle between the ground and the rod $AB$, and the \emph{caster angle}~$\lambda$ as the angle between the vertical axis and the front rod. These are the geometric parameters characterising the model. The length~$l$ and the \emph{fork lower}~$b$ are clearly related to $w$, $\f$ and $\lambda$ by simple trigonometric relations.

In order to describe the configuration of the system, we introduce the inertial fixed reference frame $\Sigma = (O;X,Y,Z)$, with origin in $O$ and $Z$-axis perpendicular to the ground and directed opposite to gravity. Let us assume that the $X$-axis passes through the two contact points when the bicycle is in the trivial configuration, whereas the $Y$-axis is chosen following the right-hand rule - this rule will be used throughout the paper. 

\begin{figure}[htb]
\begin{center}
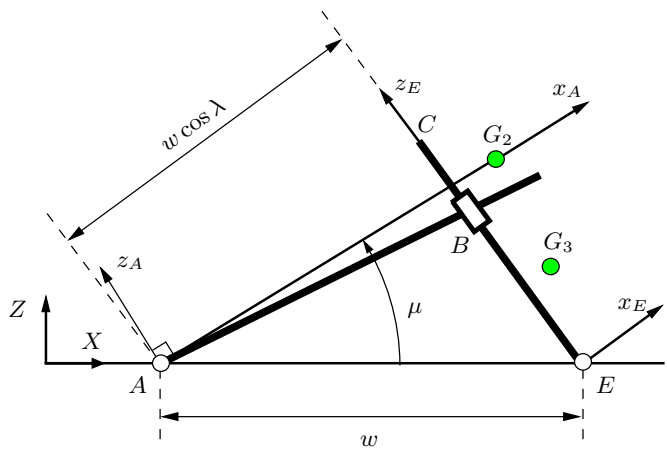
\caption{Reference frames introduced for the TMS referring to the trivial configuration. We choose $S_A$ with $x_A$ passing through the centre of mass $G_2$, and $S_E$ with $z_E$ coinciding with the steering handle. Coordinates are right-handed.}
\label{fig:TMS_frames}
\end{center}
\end{figure}

Further, we introduce a local reference frame attached to each frame. 
Referring to the trivial configuration, let~$S_A = (A;x_A, y_A, z_A)$ be centred in the rear contact point~$A$ with~$x_A$ passing through the rear frame centre of mass~$G_2$. The second reference frame $S_E = (E; x_E, y_E, z_E)$ is such that $x_E$ is normal to the front frame $BE$. Both $z_A$ and $z_E$ are pointing upwards. Note that the choice of $S_E$ is convenient because the wheelbase $w$ is constant for the TMS model, as we will show later. In Figure~\ref{fig:TMS_frames}, we draw the reference frames in the trivial configuration.

The orientations of the local reference frames with respect to $\Sigma$ are expressed by four Euler angles. We follow the choice in~\cite{caim_simai2012}. Hence, we have the \emph{yaw angle} $\theta$, taken about the vertical $Z$-direction, which identifies the angle between the $X$-axis and the line of intersection of the symmetry plane of the rear frame with the ground. The \emph{roll angle}~$\alpha$ is the angle that the rear plane makes with the normal plane to the ground. The \emph{pitch angle}~$\mu$ identifies the rotation about the~$y_A$-axis. The \emph{steering angle} $\psi$ characterises the rotation about the steering axis, which coincides with the rod~$CE$.

Note that the pitch angle $\mu$ is related to~$\f$ by a constant rotation angle. We take $\alpha$ positive for clockwise rotations, in order to have a positive angle when the bicycle leans to the left. We use the same convention also for the pitch angle, as shown in Figure~\ref{fig:TMS_frames}. Furthermore, due to physical reasons, both the roll and the steering angles have values in the interval $\ina{-\frac\pi2, \frac\pi2}$.

The direct transformations from $\Sigma$ to the local reference frames are obtained by writing the rotation matrices which characterise each rotation. In particular, we choose the \emph{alias} approach to represent these rotations, so that they change the coordinate system. The four rotation matrices are
\begin{align*}
\mc{R}_1(\theta) &=
\begin{pmatrix}
 \cos\theta & \sin\theta & 0 \\
-\sin\theta & \cos\theta & 0 \\
    0       &     0      & 1
\end{pmatrix}, &
\mc{R}_2(-\alpha) &=
\begin{pmatrix}
1 &     0      & 0 \\
0 & \cos\alpha & -\sin\alpha \\
0 & \sin\alpha &  \cos\alpha
\end{pmatrix}, \\
\mc{R}_3(-\eta) &=
\begin{pmatrix}
 \cos\eta & 0 & \sin\eta \\
    0    & 1 &     0    \\
-\sin\eta & 0 & \cos\eta
\end{pmatrix}, &
\mc{R}_4(\psi) &=
\begin{pmatrix}
 \cos\psi & \sin\psi & 0 \\
-\sin\psi & \cos\psi & 0 \\
    0     &     0    & 1
\end{pmatrix},
\end{align*}
where $\eta$ stands either for the pitch angle $\mu$ or for the caster angle~$\lambda$.  The overall rotation characteristic of each local frame is given by composing these matrices. Having chosen the alias transformations, composition of successive rotations is then obtained by multiplying matrices following the inverse order of rotations. Therefore, the rotation matrix associated to~$S_A$ is
\[
\mc{R}_A = \mc R_3(-\mu) \mc R_2(-\alpha) \mc R_1(\theta),
\]
while the rotation matrix relative to the front frame can be expressed as
\[
\mc{R}_E = \mc R_4(\psi) \mc R_3(-\lambda) \mc R_2(-\alpha) \mc R_1(\theta).
\]
This latter transformation can be also expressed by means of only three auxiliary Euler angles. Following~\cite{caim_simai2012}, we introduce the yaw angle $\wt\theta$, roll angle $\wt \alpha$, and pitch angle $\wt \mu$ relative to the front frame, which simplify the expression of the above matrix using only three angles.

Since the TMS moves on an horizontal plane, rolling without slipping and free of deviating from the vertical position, the rear point position can be described by the triple $(x, y, \theta)$, where~$(x, y)$ are the coordinates of the contact point $A$ in the $XY$ plane. Therefore, the configuration space $Q$ of the TMS is 7-dimensional, and we choose the coordinate vector $q = (\alpha, \psi, s_r, x,  y, \theta, s_f)$ to describe its motion, where~$s_r$ and $s_f$ are the rear and front displacements, respectively.

All the other coordinates introduced can be expressed with respect to $q$. From~\cite{caim_simai2012}, we know the expressions for the auxiliary angles, which depend on $\alpha$ and $\psi$, and the relations which define the pitch angle and the front contact point coordinates as functions of the parameters of the bicycle.

Let us consider the pitch angle. If $\ol\mu(t)$ is the \emph{effective} pitch angle, that is, the time dependent part of the pitch angle, then we can write $\mu(t) = \ol \mu(t) + \f$, as in~\cite{caim_simai2012}. Hence, the equation for the pitch angle becomes
\[
l \sin (\ol \mu(t) + \f) - b \cos(\ol \mu(t) + \lambda) = 0,
\]
which implies $\ol \mu(t)$ is always equal to zero. Consequently, any given rotation about $y_A$ is constant and independent of the other angles. Moreover, the distance between the rear and front contact points remains constant, hence the coordinates of the front contact point are given by the relations
\begin{equation}
\label{eq:front_position}
\begin{system}
x_f = x + w \cos\theta, \\
y_f = y + w \sin\theta.
\end{system}
\end{equation}

\begin{figure}[htb]
\setlength{\unitlength}{1cm}
\begin{center}
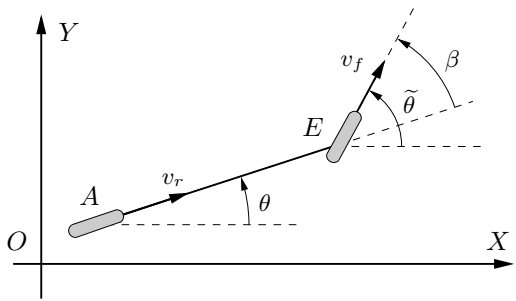
\caption{Yaw angle~$\theta$, front yaw angle $\wt\theta$ and relative yaw angle~$\beta$ in the horizontal ground plane. Angles are positive according to the right-hand rule.}
\label{fig:yaw_angles2}
\end{center}
\end{figure}

Now we consider the kinematics of the model. We only consider the nonholonomic constraints due to the rear and front contact points with the ground, while the velocities required will be derived in the following. According to~\cite{bloch:nonholo} and~\cite{monforte:nonholo}, the skate-like contact point implies that the two wheels cannot have lateral displacement and the constraints become
\begin{align}
\label{eq:rear_constraints}
&\begin{system}
\dot x \cos\theta + \dot y \sin\theta = v_r, \\
\dot x \sin\theta - \dot y \cos\theta = 0,
\end{system}
& &\text{or} &
&\begin{system}
\dot x = v_r \cos\theta, \\
\dot y = v_r \sin\theta,
\end{system}
\end{align}
for the rear contact point, and, likewise, for the front one we have
\begin{equation}
\label{eq:front_constraints}
\begin{system}
\dot x_f\cos\thetati + \dot y_f \sin\thetati = v_f, \\
\dot x_f\sin\thetati - \dot y_f \cos\thetati = 0,
\end{system}
\end{equation}
where $v_r = \dot s_r$ and $v_f = \dot s_f$. To express the second constraint with respect to the set of coordinates~$q$, we derive relation~\eqref{eq:front_position}, obtaining
\[
\begin{system}
\dot x_f = \dot x - w\dot\theta\sin\theta, \\
\dot y_f = \dot y + w\dot\theta\cos\theta,
\end{system}
\]
and substitute it in expression~\eqref{eq:front_constraints}, also using constraints~\eqref{eq:rear_constraints}. If we define the \emph{relative yaw angle}~$\beta=\beta(\alpha,\psi)=\wt\theta-\theta$ as shown in Figure~\ref{fig:yaw_angles2}, the first and second equation of~\eqref{eq:front_constraints} become
\begin{align}
\label{eq:front_constraints2a}
v_f &= \dfrac{v_r}{\cos\beta}, \\
\label{eq:front_constraints2}
\dot \theta &= \dfrac{v_r}{w} \tan\beta,
\end{align}
respectively, where
\begin{align*}
\cos\beta &= \frac{\cos\alpha\cos\psi - \sin\alpha\sin\lambda\sin\psi}{\cos\wt\alpha} & &\text{and} & \sin\beta &= \frac{\cos\lambda\sin\psi}{\cos\wt\alpha}.
\end{align*}

\begin{rmk}
The nonholonomic constraints above can also be derived from the corresponding general expressions given in~\cite{caim_simai2012}. In particular, we have to relate the angular velocities of the rear and front wheel to $v_r$ and $v_f$, respectively, and then set to zero all the parameters that vanish in the TMS model.
\end{rmk}

\section{The Lagrangian of the TMS bicycle}
\label{sec:lagrangian}

The equations of motion for a nonholonomic system may be derived in various way. We want to use the Lagrangian formalism developed in~\cite{bloch:nonholo}, which provides us with a machinery for deriving them in a clear and compact form. Indeed, this formalism generalises the Lagrange's equations of the second kind to the nonholonomic case, incorporating the nonholonomic constraints directly into the equations of motion. In order to use this approach we need the expressions of the nonholonomic contraints, which we obtained in the previous Section, and the Lagrangian of the system, given by the kinetic energy minus the potential energy. Hence, in this Section we compute it and then we introduce a symbolic formalism to handle it easily.

\subsection{Kinetic energy of the TMS bicycle}
\label{kineticenergy}

In the first part of this section we want to derive the kinetic energy of the TMS bicycle.  Theoretically, the expression of the kinetic energy of the system is simple to be obtained. In the case of a system composed of only four bodies linked between them, it is equal to the sum of the kinetic energies of each of the rigid bodies computed by means of the K\"onig's theorem. Therefore, for the TMS we have
\[
\TT = \frac12 \sum_{i=1}^{4} \left[ m_i v^2(G_i) + \scal{ \zo_i , \sigma_i(G_i) \zo_i}\right]
\]
where $\scal{\ps,\ps}$ stands for the standard scalar product, while $m_i$ is the mass of the $i$-th rigid body, $G_i$ its centre of mass, $\zo_i$ its angular velocity, and~$\sigma_i(G_i)$ represents its inertia tensor about the centre of mass with axes parallel to the local reference frame.
For instance, the kinetic energies for the rear and front wheels are
\[
 \TT_1 = \frac12 m_1 (\xp^2 + \yp^2)
\]
and
\[
 \TT_4 = \frac12 m_4 (\xp_f^2 + \yp_f^2) = \frac12 m_4 (\xp^2 + \yp^2 + w^2 \ztp^2) + m_4 w \,\ztp (-x\sin\zt + y\cos\zt),
\]
respectively, where we used the expression of the front contact point velocity given by~\eqref{eq:front_position} and recalling that both wheels have no distributed mass.

For the rear frame we note that, with respect to the local reference frame~$S_A=(A; x_A,y_A,z_A)$, the centre of mass $G_2$ has coordinates~$(G_2-A)=(l_2, 0, 0)$, as shown in Figure~\ref{fig:TMS_frames}. By using the rotation matrix~$\mathcal{R}_A$ to express the unit local vector $e_{1A}$ with respect to the inertial reference frame, we can compute $(G_2-O)_\Sigma$, and deriving this vector with respect to time we determine its velocity.
Still using the matrix~$\mathcal{R}_A$, it is also possible to obtain the expression of the angular velocity,
and after a straightforward calculation we obtain the kinetic energy
\begin{equation}
\label{eq:kin_T2}
\begin{split}
\TT_{2} &= \frac12 m_2 \left[\xp^2 +  \yp^2 + l_2^2 \zap^2 \sin^2\mu + l_2^2\ztp^2(\cos^2\mu + \sin^2\mu \sin^2 \za)\right. \\
&\quad + 2 l_2^2 \zap \ztp \sin\mu \cos\mu \cos\za + 2 l_2 \ztp \cos\mu (-\xp\sin\zt + \yp \cos\zt) \\
&\quad\left.+ 2 l_2 \zap \sin\mu \cos\za (-\xp\sin\zt + \yp \cos\zt) - 2 l_2 \ztp \sin\mu \sin\za (\xp\cos\zt + \yp \sin\zt)\right] \\
&\quad + \frac12 \left[ I_{xx2} \left(\ztp\sin\mu \cos\za - \zap\cos\mu\right)^2 + I_{yy2} \ztp^2\sin^2\varphi + I_{zz2} \left(\ztp\cos\mu \cos\za + \zap\sin\mu\right)^2 \right] \\
&\quad + I_{xz2} \left(\ztp\cos\mu \cos\za + \sin\mu \zap \right)
 \left(\sin\varphi \cos\za \,\ztp -\cos\mu \,\zap\right).
\end{split}
\end{equation}
Note that we chose $\mu$ such that the position of the centre of mass $G_2$ in $S_A$ is characterised by the length $l_2$ only, and we assumed that $I_{xz2}$ is the only nonzero product of inertia of the tensor~$\sigma_2(G_2)$.

Likewise, we can compute the kinetic energy of the front frame. However, it has even a longer expression and we skip this additional computation. We notice that, despite the conceptual simplicity of the calculations, the expression of the kinetic energy cannot be easily handled to obtain the equations of motion. Therefore, we need to introduce a different approach to deal with the kinetic energy of the bicycle.


\subsection{Symbolic form of the equations of motion}

In order to obtain the equations of motion of the system we want to follow~\cite{bloch:nonholo}, hence we need an expression of the Lagrangian. We decided to use a semi-analytic versions of both the kinetic and the potential energies. This procedure results very useful to write the equations of motion and solve them numerically, understanding the qualitative behaviour of the system.

Therefore, we write the kinetic energy in a symbolic form, where each coefficient of the velocity terms is defined as a nonlinear function  at most of  the generalised coordinates. These nonlinear functions are closely related to the Hessian of the kinetic energy itself. In particular, we choose to write the symbolic form of the kinetic energy as the positive definite quadratic function
\begin{equation}
\begin{split}
\label{eq:kinetic_free}
\TT(\alpha, \psi, \theta, \dot\alpha, \dot\psi, \dot\theta, \dot x, \dot y) 
&= \frac{1}{2} M(\dot x^2 + \dot y^2) + \frac{1}{2} a(\psi) \dot\alpha^2  + \frac{1}{2} b(\alpha, \psi) \dot\theta^2 + \frac{1}{2} P \dot\psi^2 \\
&\quad +  c(\alpha, \psi) \dot\alpha \dot\theta + d(\psi)  \dot\alpha \dot\psi +  e(\alpha, \psi) \dot\theta \dot\psi \\
&\quad + f(\za,\psi) (-\dot x \sin\theta + \dot y \cos\theta)\ztp + k(\za, \psi) (-\xp \sin\zt + \dot y \cos\theta) \zap \\
&\quad +  h(\alpha, \psi) (-\xp \sin\zt + \dot y \cos\theta) \dot\psi + l(\alpha, \psi) (\dot x \cos\theta + \dot y \sin\theta)\ztp \\
&\quad + m(\psi)  (\dot x \cos\theta + \dot y \sin\theta) \dot\psi,
\end{split}
\end{equation}
where each function can be computed symbolically with a computer-assisted method. The dependence of each function on the generalised coordinates is given in the argument of the functions. We choose a symbolic expression in which the functions depend at most on the roll and steering angle, and the yaw angle~$\theta$ always appears in expressions which remind the nonholonomic constraints for the rear wheel. This will be very convenient when the nonholonomic constraints will be introduced.

\begin{rmk}
In this expression of the kinetic energy we only assumed that the centres of mass~$G_2$ and~$G_3$ lie in the rear and front planes, respectively, as usually a bicycle is symmetric with respect to these planes. No other assumptions are required.
\end{rmk}

Similarly, we assume that the potential energy is a function of the roll and steering angles, that is,~$V = V(\alpha, \psi)$, since the conservative forces acting on the system naturally depend only on these coordinates. For instance, if the system is only subjected to the gravity force, we have
\begin{equation}
\label{eq:potential}
V(\alpha, \psi) 
= g u(\alpha, \psi).
\end{equation}
where
\begin{equation}
\label{eq:fnc_u}
u(\alpha, \psi) = m_2 l_2 \sin\mu \cos\alpha + m_3\left[h_3\cos\lambda\cos\alpha + l_3(\sin\lambda\cos\alpha\cos\psi - \sin\alpha\sin\psi)\right].
\end{equation}

Now we can write the classical Lagrangian $L(q, \dot q) = \TT(q, \dot q) - V(q)$ and derive the equations of motion. This will be presented in detail in the next section.

\section{Dynamics and equations of motion}
\label{sec:motion}

Following the approach in~\cite{bloch:nonholo}, on the configuration space $Q$ we choose the bundle~$Q \to R$, where the base~$R=S^1 \times S^1 \times \R$ is parametrized by $(\alpha, \psi, s_r)$ and $(\alpha, \psi, s_r, x, y, \theta, s_f) \mapsto (\alpha, \psi, s_r)$ is the trivial projection to~$R$. To keep the exposition simple, we do not consider the symmetry of the problem. Being $q=(r,s)$, let $\dot r = (\dot\alpha, \dot\psi, v_r)$ and $\dot s = (\dot x, \dot y, \dot\theta, v_f)$ be the \emph{base} and the \emph{fibre} velocities, respectively, where $\dot s_r = v_r$ and $\dot s_f = v_f$. Then, the constraints can be written as~$\dot s^a = - A_i^a \dot r^i$, where~$A(q)$ represents the Ehresmann connection. See~\cite{bloch:nonholo} and~\cite{monforte:nonholo} for details. Hence, the equations of motion are
\begin{equation}
\label{eq:euler_lagrange}
\frac{d}{dt}\frac{\dpe L_c}{\dpe \dot r^i} - \frac{\dpe L_c}{\dpe r^i} + A_i^a \frac{\dpe L_c}{\dpe s^a} = - \frac{\dpe L}{\dpe \dot s^b} B^b_{ij} \dot r^j,
\end{equation}
where
\begin{equation}
\label{eq:curvature_coefficients}
B^b_{ij} = \frac{\dpe A_i^b}{\dpe r^j} - \frac{\dpe A_j^b}{\dpe r^i} + A_i^a \frac{\dpe A_j^b}{\dpe s^a} - A_j^a \frac{\dpe A_i^b}{\dpe s^a},
\end{equation}
are the coefficients of the curvature, 
while $L_c(r^i, s^a, \dot r^i) = L(r^i, s^a, \dot r^i, -A_i^a(r, s)\dot r^i)$ is the \emph{constrained Lagrangian}, obtained by introducing the nonholonomic constraints in the Lagrangian $L(r^i, s^a, \dot r^i,\dot s^a)$. Note that $i,j = 1,2, 3$ and $a,b= 1,2,3,4$. From the constraints~\eqref{eq:rear_constraints},~\eqref{eq:front_constraints2a} and~\eqref{eq:front_constraints2} one can read off the components of the Ehresmann connection, namely,
\begin{align*}
A_3^1 &= -\cos\theta, & A_3^2 &= -\sin\theta, & A_3^3 &= -\frac{1}{w}\tan\beta, & A_3^4 &= - \frac{1}{\cos\beta},
\end{align*}
while the remaining $A_i^a$ are zero. Consequently, the nonzero curvature coefficients are
\begin{align*}
B_{13}^3 &= - B_{31}^3 = - \frac{\dpe}{\dpe \alpha} A_3^3 = \frac{1}{w}\frac{\dpe}{\dpe\alpha}\tan\beta = \frac{\dpe z(\alpha, \psi)}{\dpe \alpha}, \\
B_{23}^3 &= - B_{32}^3 = - \frac{\dpe}{\dpe \psi} A_3^3 = \frac{1}{w}\frac{\dpe}{\dpe\psi}\tan\beta = \frac{\dpe z(\alpha, \psi)}{\dpe \psi}, \\
B_{13}^4 &= - B_{31}^4 = - \frac{\dpe}{\dpe \alpha} A_3^4 = \frac{\dpe}{\dpe\alpha}\frac{1}{\cos\beta}, \\
B_{23}^4 &= - B_{32}^4 = - \frac{\dpe}{\dpe \psi} A_3^4 = \frac{\dpe}{\dpe\psi}\frac{1}{\cos\beta},
\end{align*}
where
\begin{equation}
\label{eq:fnc_z}
z(\alpha, \psi) = \frac{\tan\beta}{w} = \frac{\cos\lambda\sin\psi}{w(\cos\alpha\cos\psi - \sin\alpha\sin\lambda\sin\psi)}.
\end{equation}
Therefore, since~$s^a$ are cyclic variables, that is, the constrained Lagrangian~$L_c$ does not depend on the~$s^a$'s, the equations of motion are
\begin{align}
&\frac{d}{d t}\frac{\dpe L_c}{\dpe \dot\alpha} - \frac{\dpe L_c}{\dpe\alpha} = - \left.\frac{\dpe L}{\dpe\dot\theta}\right|_c B_{13}^3 v_r, \label{eq:alpha} \\
&\frac{d}{d t}\frac{\dpe L_c}{\dpe \dot\psi} - \frac{\dpe L_c}{\dpe\psi} = - \left.\frac{\dpe L}{\dpe\dot\theta}\right|_c B_{23}^3 v_r, \label{eq:psi} \\
&\frac{d}{d t}\frac{\dpe L_c}{\dpe v_r} = - \left.\frac{\dpe L}{\dpe\dot\theta}\right|_c (B_{31}^3 \dot\alpha + B_{32}^3 \dot\psi). \label{eq:v_r}
\end{align}
We remark that the constrains are substituted into the right-hand side \emph{after} the partial derivative of~$L$ with respect to~$\dot\theta$ is taken.
 
We now explicit the three equations of motions \eqref{eq:alpha}, \eqref{eq:psi} and~\eqref{eq:v_r}, and by using the nonlinear functions in our symbolic expressions we easily simplify the expressions. In particular, we express symbolically the constrained Lagrangian as
\[
\begin{split}
L_c(\alpha, \psi, \dot\alpha, \dot\psi, v_r) &= \frac{1}{2} v_r^2 n(\alpha, \psi) + \frac{1}{2} \dot\alpha^2 a(\psi) + \frac{1}{2} P \dot\psi^2 + \\
&\quad + v_r \dot\alpha p(\alpha, \psi) + \dot\alpha\dot\psi d(\psi) + \dot\psi v_r q(\alpha, \psi) - u(\alpha, \psi),
\end{split}
\]
where we introduce the additional nonlinear functions
\begin{equation}
\label{eq:fncs_new}
\begin{system}
n(\alpha, \psi) = M + z^2(\alpha, \psi) b(\alpha, \psi) + 2 z(\alpha, \psi)l(\alpha, \psi), \\
p(\alpha, \psi) = z(\alpha, \psi)c(\alpha, \psi), \\
q(\alpha, \psi) = z(\alpha, \psi)e(\alpha, \psi) + m(\psi).
\end{system}
\end{equation}
Then, we can substitute this expression in the equations of motion, and after a lengthy but straightforward calculation, equation~\eqref{eq:alpha} becomes
\begin{equation}
\label{eq:alpha_expl}
\begin{split}
&\ddot \alpha a(\psi) + \ddot \psi d(\psi) + \dot v_r p(\alpha, \psi)+ \dot\alpha \dot\psi \frac{\dpe a(\psi)}{\dpe\psi} + \dot\alpha v_r \frac{\dpe z(\alpha, \psi)}{\dpe\alpha}c(\alpha, \psi) + \dot\psi^2 \frac{\dpe d(\psi)}{\dpe\psi} + \\
&\quad + \dot\psi v_r \left[\frac{\dpe z(\alpha, \psi)}{\dpe\psi}c(\alpha, \psi) + z(\alpha, \psi) \frac{\dpe c(\alpha, \psi)}{\dpe\psi}- z(\alpha, \psi) \frac{\dpe e(\alpha, \psi)}{\dpe\alpha}\right]+ \\
&\quad - v_r^2 \left[\frac{1}{2}z(\alpha, \psi) \frac{\dpe b(\alpha, \psi)}{\dpe \alpha} + \frac{\dpe l(\alpha, \psi)}{\dpe\alpha}\right]z(\alpha, \psi) + \frac{\dpe V(\alpha, \psi)}{\dpe \alpha} = 0.
\end{split}
\end{equation}
Similarly, from equation~\eqref{eq:psi} we obtain
\begin{equation}
\label{eq:psi_expl}
\begin{split}
&\ddot \alpha d(\psi) + \ddot \psi P + \dot v_r q(\alpha, \psi) + \dot\psi v_r \frac{\dpe z(\alpha, \psi)}{\dpe\psi}e(\alpha, \psi) - \frac{1}{2}\dot\alpha^2 \frac{\dpe a(\psi)}{\dpe\psi} + \\
&\quad + \dot\alpha v_r \left[\frac{\dpe z(\alpha, \psi)}{\dpe\alpha}e(\alpha, \psi) + z(\alpha, \psi) \frac{\dpe e(\alpha, \psi)}{\dpe\alpha}- z(\alpha, \psi) \frac{\dpe c(\alpha, \psi)}{\dpe\psi}\right]+ \\
&\quad - v_r^2 \left[\frac{1}{2}z(\alpha, \psi) \frac{\dpe b(\alpha, \psi)}{\dpe \psi} + \frac{\dpe l(\alpha, \psi)}{\dpe\psi}\right]z(\alpha, \psi) + \frac{\dpe V(\alpha, \psi)}{\dpe \psi} = 0,
\end{split}
\end{equation}
and~\eqref{eq:v_r} turns into
\begin{equation}
\label{eq:vr_expl}
\begin{split}
&\ddot \alpha p(\alpha, \psi) + \ddot \psi q(\alpha, \psi) + \dot v_r n(\alpha, \psi) + \dot\alpha^2 z(\alpha, \psi) \frac{\dpe c(\alpha, \psi)}{\dpe \alpha} + \\
&\quad + \dot\alpha\dot\psi \left[\frac{\dpe c(\alpha, \psi)}{\dpe \psi} + \frac{\dpe e(\alpha, \psi)}{\dpe \alpha}\right]z(\alpha, \psi)+ \dot\psi^2 \left[z(\alpha, \psi)\frac{\dpe e(\alpha, \psi)}{\dpe\psi} + \frac{\dpe m(\psi)}{\dpe\psi}\right] + \\
&\quad + \dot\alpha v_r \left[\left(z(\alpha, \psi)b(\alpha, \psi) + l(\alpha, \psi)\right) \frac{\dpe z(\alpha, \psi)}{\dpe\alpha} + \left(z(\alpha, \psi) \frac{\dpe b(\alpha, \psi)}{\dpe\alpha} + 2 \frac{\dpe l(\alpha, \psi)}{\dpe\alpha}\right)z(\alpha, \psi)\right] + \\
&\quad + \dot\psi v_r \left[\left(z(\alpha, \psi)b(\alpha, \psi) + l(\alpha, \psi)\right) \frac{\dpe z(\alpha, \psi)}{\dpe\psi} + \left(z(\alpha, \psi) \frac{\dpe b(\alpha, \psi)}{\dpe\psi} + 2 \frac{\dpe l(\alpha, \psi)}{\dpe\psi}\right)z(\alpha, \psi)\right] = 0.
\end{split}
\end{equation}
All the terms in these equations are known, and the nonlinear functions allowed us
to simplify remarkably the expressions.

\begin{rmk}
The nonlinear functions can determined as follows. Referring to expression~\eqref{eq:kinetic_free}, each nonlinear function can be determined through symbolic calculus by taking the partial second derivatives of the kinetic energy with respect to the speeds which multiplies the nonlinear function itself. However, for those functions which multiply also functions of $\theta$, we evaluate the second partial derivative either at~$\theta=0$ or at~$\theta=\pi/2$ in order to make the function of $\theta$ equal to one.
This procedure can be done with any computational software program based on symbolic calculus. We used Wolfram Mathematica. Then, we assign the geometric parameters in the program, and each function is a variable which depends on the roll and the steering angles at most. The remaining nonlinear functions are defined through relations~\eqref{eq:fnc_z} and~\eqref{eq:fncs_new}.
\end{rmk}

Note that the differential equations above for the base coordinates $r=(\alpha,\psi,s_r)$ give, solved together with the nonholonomic constraints, the dynamics of the system. General solutions for this set of differential equations can be obtained through numerical integration. However, there are two special solutions which can be written in closed form. These solutions form two classes particularly important for studying the stability of the system, and we discuss them in the following.

\subsection{Rectilinear motions}
\label{sct:rect}

The first class of special solutions is given by \emph{rectilinear motions} of the TMS in the upward position, for which
\[
q_0(t) = (0, 0, v_0 t, v_0 t \cos \theta_0, v_0 t \sin \theta_0, \theta_0, v_0 t),
\]
where $v_0\in \R$ and $\theta_0\in \inca{0,2\pi}$ are arbitrary chosen. It can be checked that the equations of motion and the nonholonomic constraints are satisfied by any solution of this form. Indeed, the partial derivatives of the potential energy $V(\alpha,\psi)$, as well as the other nonlinear functions and their partial derivatives, are equal to zero when evaluated at~$(\alpha_0, \psi_0)=(0,0)$. The only parameters which characterise the rectilinear motion are the initial rear speed~$v_0$ and the yaw angle~$\theta_0$.

Although the rectilinear solution above exists for any choice of the geometric parameters characterising the model, its stability depends on the geometry of the system and on the rear speed~$v_0$. In particular, we want to see when this solution is \emph{asymptotically stable}, that is, both the roll and the steering angles return to zero whenever perturbed from the rectilinear solution.

We determine the stability of this solution by linearising the system about the solution and then studying the eigenvalues of the Jacobian matrix obtained. Hence, we need to write the equations of motion as a system of first-order differential equations. Since the rear speed is constant by definition of rectilinear solution, that is, $v_r(t) = v_0$, we can write the equation for $\alpha$ and $\psi$ only. Therefore, let~$\xi= (\alpha, \dot\alpha, \psi, \dot\psi)$, then $\dot \xi = F(\xi)$ defines the dynamics of these angles. Note that $F(\xi)$ depends on the equations~\eqref{eq:alpha_expl} and~\eqref{eq:psi_expl}, on the rear speed $v_0$ and on the geometric parameters of the TMS. Linearising the system about~$\xi_0=(0,0,0,0)$, we have $\dot \xi = J(\xi_0)\xi$, where~$J(\xi)$ is the Jacobian of the nonlinear function~$F(\xi)$.

\begin{table}[tb]
\caption{Nonzero geometric parameters of the theoretical TMS as in~\cite{meijaard:science}. The coordinates of the two centres of mass are expressed in the trivial configuration with respect to the fixed reference frame $\Sigma$ with origin in $A$. We chose $Z$ to be directed upward, thus $z_2$ and $z_3$ have signs opposite to those in the original paper.}
\label{tab:values1}
\centering
    \begin{tabular}{lll}
    \toprule
    \textbf{Symbol} & \textbf{Parameter} & \textbf{Value} \\
    \midrule
    $w$ & wheelbase & \SI{1}{m} \\
    \midrule
    $\lambda$ & caster angle & \SI{5}{\degree} \\
    \midrule
    $m_2$ & rear frame mass & \SI{10}{\kilo\gram} \\
    \midrule
    $(x_2, z_2)$ & rear frame centre of mass & (\SI{1.2}{m}, \SI{0.4}{m}) \\
    \midrule
    $m_3$ & front frame mass & \SI{1}{\kilo\gram} \\
    \midrule
    $(x_3, z_3)$ & front frame centre of mass & (\SI{1.02}{m}, \SI{0.2}{m}) \\
    \bottomrule
    \end{tabular}
\end{table}

In order to study the eigenvalues of $J(\xi_0)$, we refer to the geometric parameters of the theoretical TMS given in~\cite{ meijaard:science}. We listed these parameters in Table~\ref{tab:values1} referring to the trivial configuration. Note that, by simple trigonometric relations, we have
\begin{align}
\label{eq:trig_rel}
\mu &= \arctan\left(\frac{z_2}{x_2}\right), &  l_2 &= \frac{x_2}{\cos\mu},  \\
 l_3 &= z_3\sin\lambda - (w - x_3)\cos\lambda, & h_3 &= z_3 \cos\lambda +(w-x_3)\sin\lambda.
\end{align}

\begin{figure}[htb]
\centering
\includegraphics[width=0.65\textwidth]{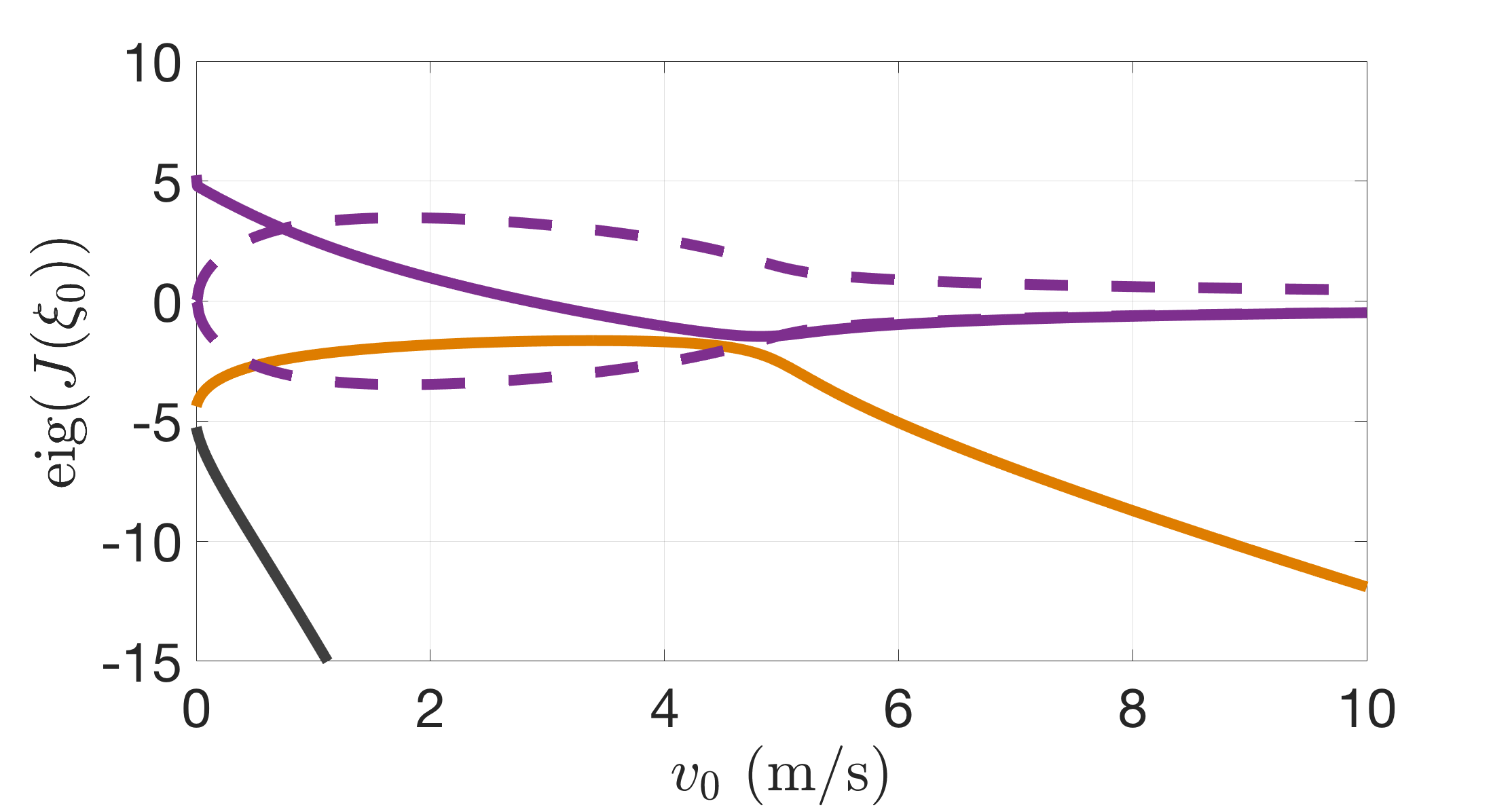}
\caption{Dependence of the four eigenvalues of the Jacobian $J(\xi_0)$ on the rear speed $v_0$ for the theoretical TMS. Real (solid line) and imaginary (dashed line) parts of the eigenvalues are plotted. There are two real eigenvalues and a pair of complex conjugate eigenvalues. The real eigenvalues are negative for every value of the rear speed, while the complex conjugate pair has negative real part only if~$v_0\ge\SI{2.85}{\meter\per\second}$. }
\label{fig:eigen_rect}
\end{figure}

Figure~\ref{fig:eigen_rect} shows the real and imaginary parts of the four eigenvalues of the Jacobian numerically computed for increasing value of the rear speed~$v_0$. We have two real eigenvalues which are negative for every value of the rear speed, and a pair of complex conjugate eigenvalues whose common real part becomes negative when $v_0$ is larger than a critical value $v_\mr{cr} \simeq \SI{2.85}{\meter\per\second}$. Hence, from the numerics we observe that the rectilinear solution is asymptotically stable only if $v_0\ge v_\mr{cr}$, that is, when all the eigenvalues have negative real part. 

\begin{rmk}
Note that, as the eigenvalues of $J(\xi_0)$ characterise the linearised system, the critical value $v_\mr{cr}$ obtained above is the same as the one in~\cite{ meijaard:science}, in which the author considered a linear model. However, we will numerically show that our model presents a wider range of stability.
\end{rmk}

\subsection{Circular motions}

The second class of solutions is given by \emph{circular motions} of the TMS characterised by constant roll and steering angles, as well as constant rear speed, namely,
\[
r_c(t) = (\alpha_c, \psi_c, v_c t).
\]
Note that this class of solutions generalises the one of rectilinear motions. By substituting the solution~$r_c(t)$ in the equations of motion~\eqref{eq:alpha_expl}, \eqref{eq:psi_expl} and~\eqref{eq:vr_expl}, we observe that the latter is always satisfied, while from the first couple of equations we obtain the system
\begin{equation}
\label{eq:alg_system}
\begin{system}
v_c^2 \left[\dfrac{1}{2}z(\alpha_c, \psi_c) \left.\dfrac{\dpe b(\alpha, \psi)}{\dpe \alpha}\right|_{(\alpha_c, \psi_c)} + \left.\dfrac{\dpe l(\alpha, \psi)}{\dpe\alpha}\right|_{(\alpha_c, \psi_c)}\right]z(\alpha_c, \psi_c) - \left.\dfrac{\dpe V(\alpha, \psi)}{\dpe \alpha}\right|_{(\alpha_c, \psi_c)} = 0, \\
\\
v_c^2 \left[\dfrac{1}{2}z(\alpha_c, \psi_c) \left.\dfrac{\dpe b(\alpha, \psi)}{\dpe \psi}\right|_{(\alpha_c, \psi_c)} + \left.\dfrac{\dpe l(\alpha, \psi)}{\dpe\psi}\right|_{(\alpha_c, \psi_c)}\right]z(\alpha_c, \psi_c) - \left.\dfrac{\dpe V(\alpha, \psi)}{\dpe \psi}\right|_{(\alpha_c, \psi_c)} = 0,
\end{system}
\end{equation}
where the nonlinear functions are those defined before. Therefore, the circular motion~$r_c(t)$ exists whenever the triple $(\alpha_c, \psi_c, v_c)$ satisfies this system of algebraic equations~\eqref{eq:alg_system}. The rectilinear solution, considered as particular circular motion, always satisfies this system.

Hence, we want to see if there exist nontrivial solutions which satisfy the system. In particular, fixed a value among $\alpha_c$, $\psi_c$ and $v_c$, we determine the remaining two by solving system~\eqref{eq:alg_system}. 
If a nontrivial solution exists, then from the constraint on the yaw angle we have $\dot \theta = v_c z(\alpha_c, \psi_c)$, which implies $\theta(t) = \theta_c t + \theta_0$, where $\theta_c=v_c z(\alpha_c, \psi_c)$ and $\theta_0\in \inca{0,2\pi}$ arbitrary. By using also the other nonholonomic constraints, we obtain
\[
q_c(t) = \left(\alpha_c, \psi_c, v_c t, \frac{w}{\tan\beta_c}\sin(\theta_c t + \theta_0), -\frac{w}{\tan\beta_c}\cos(\theta_c t + \theta_0), \theta_c t + \theta_0, \frac{v_c}{\cos\beta_c} t\right),
\]
with $\beta_c$ the relative yaw angle evaluated at $(\alpha_c, \psi_c)$. Note that the rear contact point describes a circle of radius $R = w/(\tan\beta_c)$ on the ground plane.

\begin{rmk}
Note that nontrivial circular solutions are closely related to the nonlinearity of our model. Indeed, if we consider a linear model as in~\cite{meijaard:science}, then~\eqref{eq:alg_system} becomes a linear system. Because the coefficients of the variables in each of the two equations are independent, the only possible solution is the trivial one, which characterises to rectilinear motions of the TMS.
\end{rmk}

\subsubsection{Existence of circular motions}

\begin{figure}[htb]
    \centering
    \begin{subfigure}[b]{0.45\textwidth}
        \includegraphics[width=\textwidth]{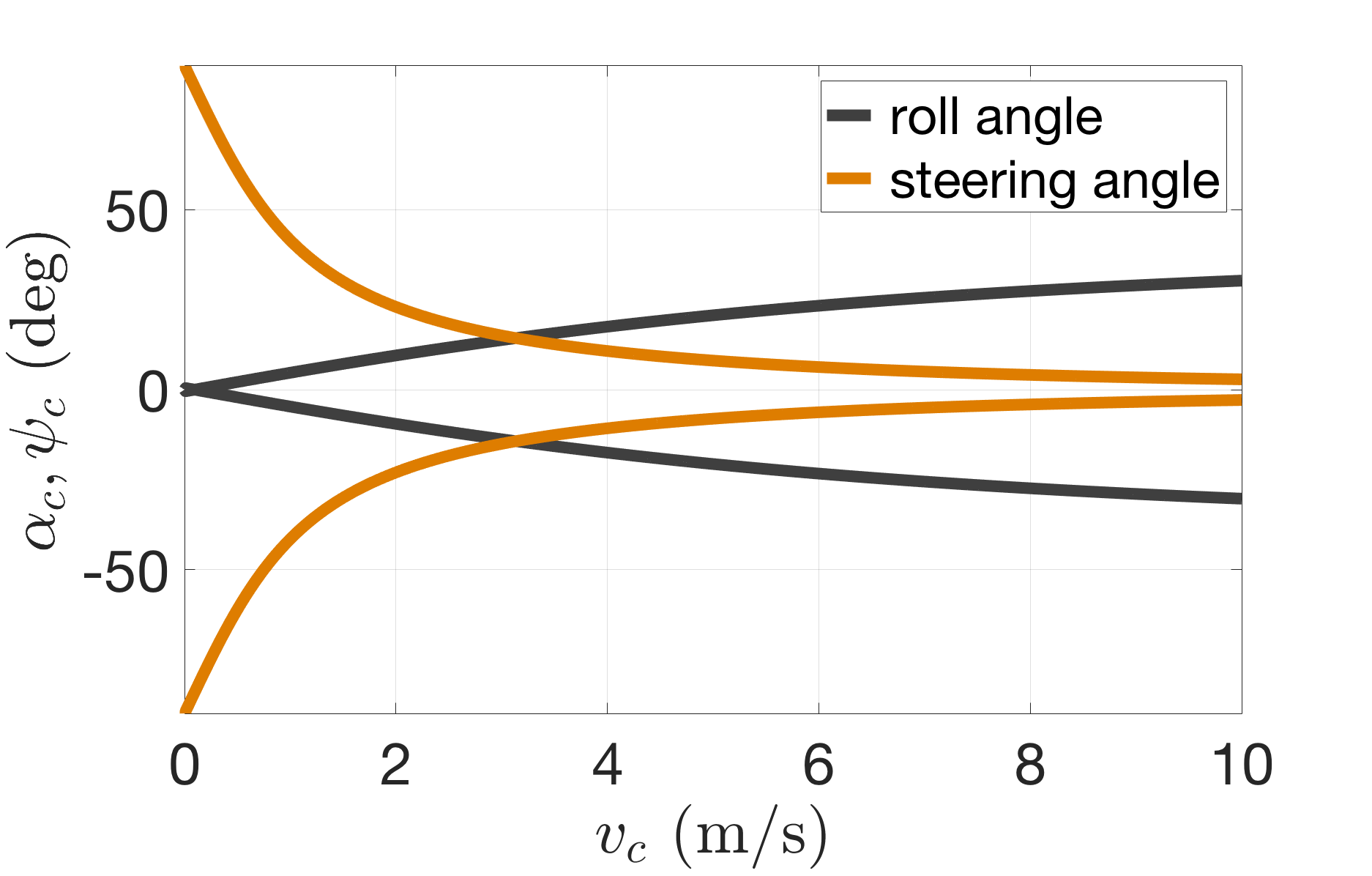}
        \caption{Caster angle $\lambda=\SI{5}{\degree}$.}\label{fig:circular_TMS}
    \end{subfigure}
    ~
    \begin{subfigure}[b]{0.45\textwidth}
        \includegraphics[width=\textwidth]{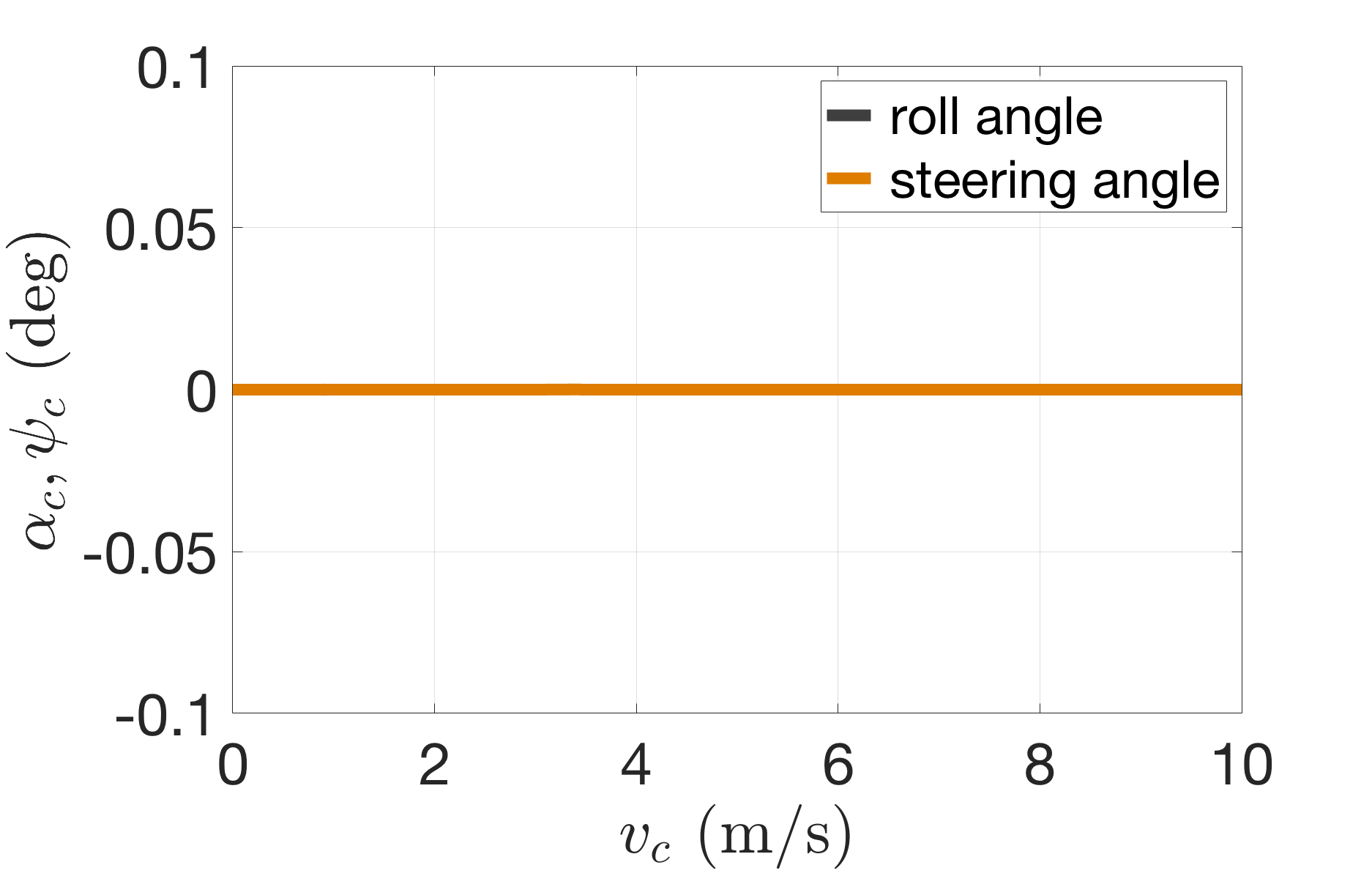}
        \caption{Caster angle $\lambda=\SI{0}{\degree}$.}\label{fig:no_circular_TMS}
    \end{subfigure}
\caption{Dependence of the roots $\alpha_c$ and $\psi_c$ for system~\eqref{eq:alg_system} on the rear speed~$v_c$. Given a value of the velocity, we have a positive and a negative pair of solutions which are symmetric, which correspond to a circular motion in counter-clockwise and clockwise direction, respectively. The existence of a circular motion strongly depends on the geometric parameters of the TMS. In this case, we chose~$v_c = \SI{3.5}{\meter\per\second}$ and the geometric parameter, apart from the caster angle, are those in Table~\ref{tab:values1}.}
\end{figure}

Before studying the stability of this class of solutions, we want to see how the existence of a circular motion depends on the geometric parameters. We choose again the geometric parameters in Table~\ref{tab:values1}. In Figure~\ref{fig:circular_TMS} we plot the roots $\alpha_c$ and $\psi_c$ of~\eqref{eq:alg_system} for different values of the rear speed.

First we note that, if the rear speed tends to zero, the roll angle goes to zero and the steering angle approaches the right angle, that is, in the limit the circular motion is realised for any front speed $v_f$ by keeping the rear frame upward ($\alpha_c=\SI{0}{\degree}$) and the handle steered by~$\psi_c=\SI{\pm90}{\degree}$. This configuration is similar to the pivoting about the fixed rear contact point described in~\cite{basu:circular} for a benchmark bicycle.

Secondly, we observe that the solution of system~\eqref{eq:alg_system} consists of two pair of symmetric roots for the roll and steering angle, namely $(\alpha_c, \psi_c)$ and $(-\alpha_c, -\psi_c)$, corresponding to one circular motion in counter-clockwise and one in clockwise direction, respectively. This is a consequence of the system symmetry and the homogeneity of the ground plane.

Then, we change the caster angle $\lambda$ to zero, while all the remaining geometric parameters are kept the same. In this case, system~\eqref{eq:alg_system} does not admit any solution except for the trivial one corresponding to the rectilinear motion. 
Therefore, we see from the numerics that the existence of nontrivial circular solutions depends on the geometric parameters of the model.


\subsubsection{Stability analysis of circular motion}

Since the rear speed $v_c$ is constant, we can study the stability by considering the system~$\dot\xi=F(\xi)$, where~$\xi=(\alpha,\psi,\dot\alpha,\dot\psi)$. Linearising along the solution~$r_c(t)$, we obtain~$\dot\xi=J(\xi_c^\pm)\xi$, where~$\xi_c^\pm=(\pm\alpha_c, \pm\psi_c, 0, 0)$ represents the pair of circular motions. Then, we compute the eigenvalues of the Jacobian matrix~$J(\xi_c^\pm)$. Note that, by symmetry, the eigenvalues are the same either we choose the positive or the negative pair.

\begin{figure}[htb]
\centering
\includegraphics[width=0.65\textwidth]{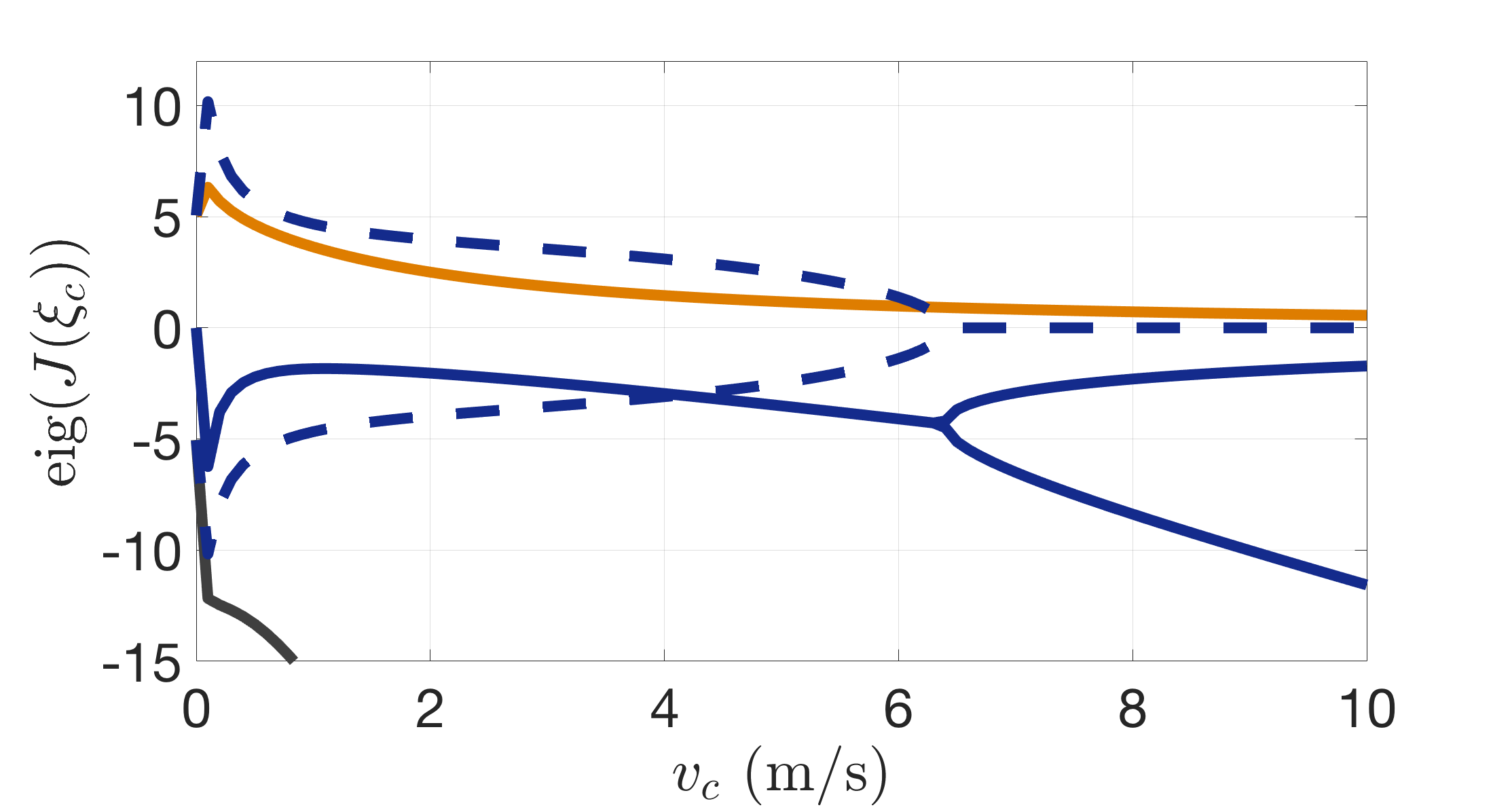}
\caption{Dependence of the four eigenvalues of the Jacobian $J(\xi_c)$ on the rear speed $v_c$ for the theoretical TMS. Real (solid line) and imaginary (dashed line) part of each eigenvalue is plotted. There is one eigenvalue whose real part is always positive, hence the circular solution is always unstable. Moreover, for $v_c\ge\SI{6.4}{\meter\per\second}$ all the eigenvalues are real.}
\label{fig:eigen_circ}
\end{figure}

Taking the geometric parameters as in Table~\ref{tab:values1}, in Figure~\ref{fig:eigen_circ} we plotted the eigenvalues of the Jacobian~$J(\xi_c^\pm)$. We observe that for any value of the rear speed at least one eigenvalue has positive real part. Hence, the numerics show that the circular solution is always unstable for these geometric parameters. This unstable behaviour of the circular solution is also described in~\cite{basu:circular}, where the authors reported that circular motions for a benchmark bicycle are always unstable when the handlebar is directed in the forward direction, as in our case.

\subsubsection{Relations between circular and rectilinear motions}

Let us consider the four dimensional phase space parametrised by~$\xi=(\alpha,\psi,\dot\alpha,\dot\psi)$. We note that the rectilinear and the circular solutions can be viewed as fixed point in this phase space. In particular, if we write the system $\dot \xi = F(\xi)$, these points are obtained as solutions of~$F(\xi)=0$, which coincides with system~\eqref{eq:alg_system}.

We observe that the origin of this phase space corresponds to the rectilinear solution~$\xi_0$, and that the circular motion corresponds to a pair fixed points located in $(\alpha_c, \psi_c, 0, 0)$ and $(-\alpha_c, -\psi_c, 0, 0)$, respectively. While the former fixed point always exists, the existence of the nontrivial pair appears to depend on the choice of the geometric parameters, as shown numerically before. Moreover, the stability of the fixed point at the origin depends on the value of the rear speed, while the fixed points corresponding to the circular motion seem to be always unstable.

\begin{figure}[htb]
    \centering
    \begin{subfigure}[b]{0.45\textwidth}
        \includegraphics[width=\textwidth]{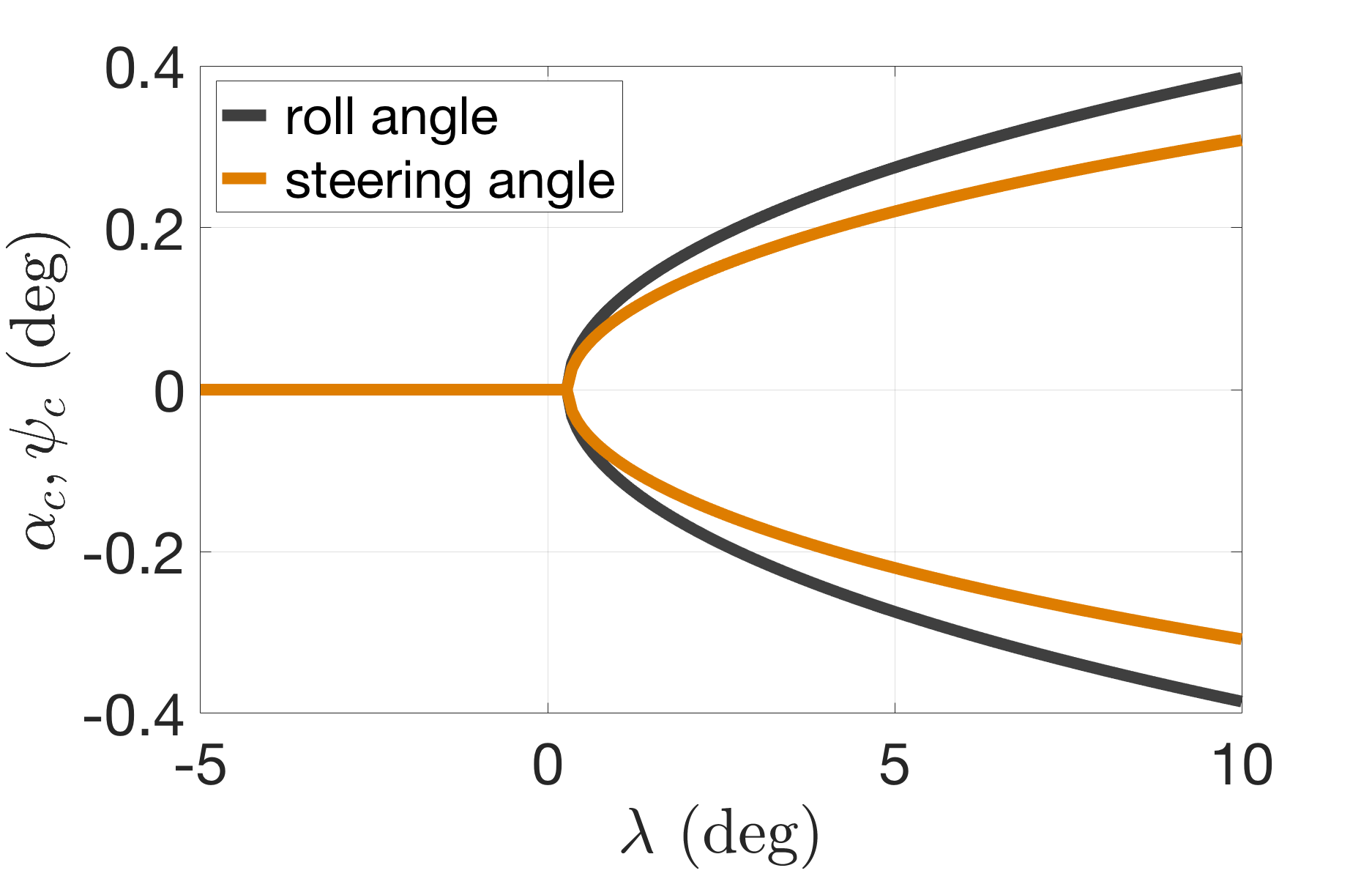}
        \caption{Variation of the caster angle $\lambda$.}\label{fig:pitch_lambda}
    \end{subfigure}
    ~
    \begin{subfigure}[b]{0.45\textwidth}
        \includegraphics[width=\textwidth]{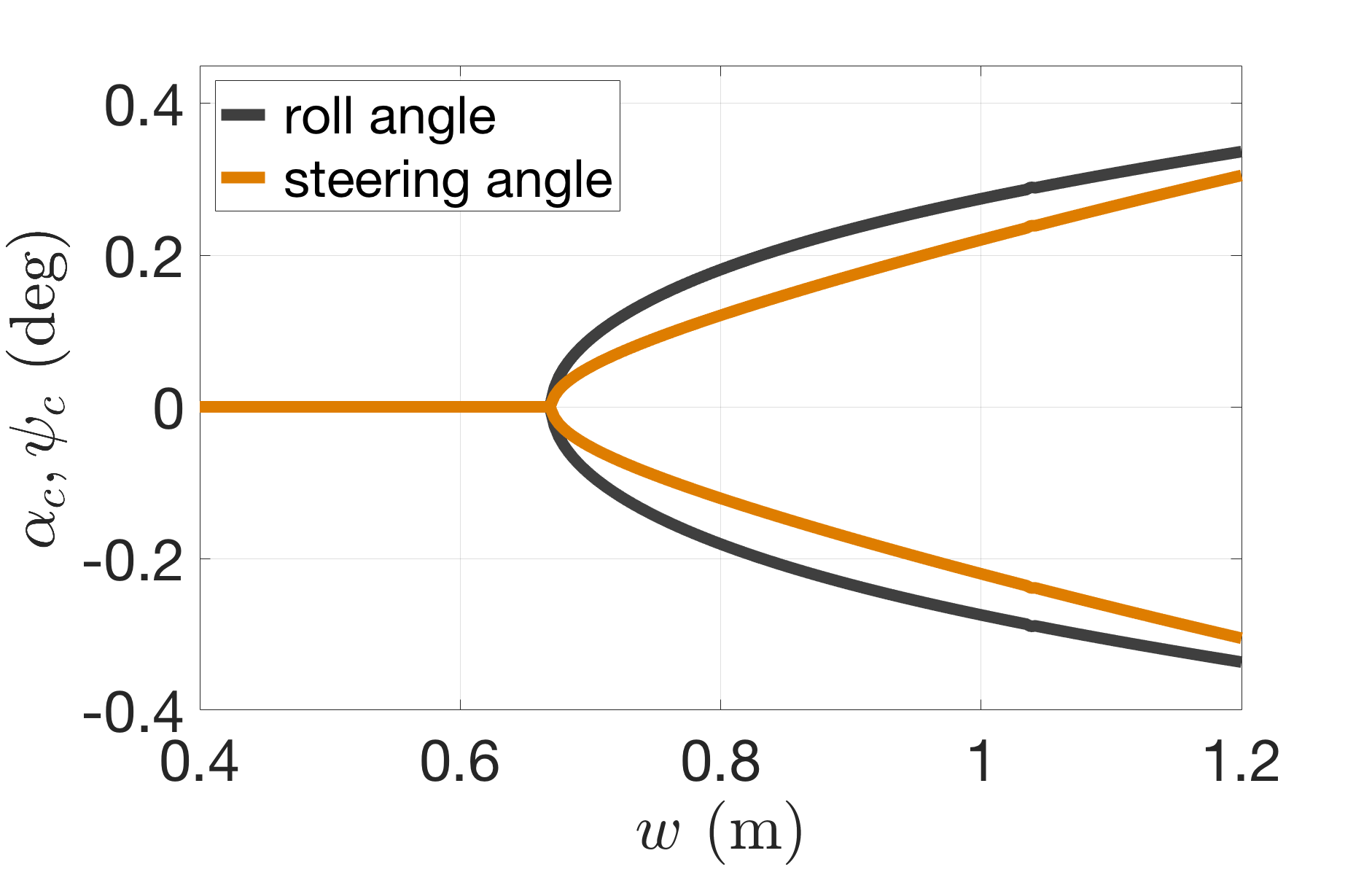}
        \caption{Variation of the wheelbase $w$.}\label{fig:pitch_w}
    \end{subfigure}
\caption{Dependence of the roots $\alpha_c$ and $\psi_c$ for system~\eqref{eq:alg_system} on the caster angle~$\lambda$ and the wheelbase~$w$, with the other geometric parameters as in Table~\ref{tab:values1} and taking $v_r = \SI{3.5}{\meter\per\second}$. For the critical value~$\lambda_\mr{cr} \simeq \SI{0.2}{\degree}$ and for the critical wheelbase $w_\mr{cr}\simeq \SI{0.66}{\meter}$, respectively, we observe a bifurcation. For any bifurcation parameter chosen, its critical value does not depend on the rear speed $v_c$.}
\label{fig:pitch}
\end{figure}

Then, let us consider the dependence of the roots $(\alpha_c, \psi_c)$ of system~\eqref{eq:alg_system} on the geometric parameters of the TMS. starting with the parameter in Table~\ref{tab:values1}, we vary the caster angle~$\lambda$ and the wheelbase~$w$, separately. In Figure~\ref{fig:pitch} we plotted the roots. From the numerics, we can determine the critical values~$\lambda_\mr{cr} \simeq \SI{0.2}{\degree}$ and $w_\mr{cr}\simeq \SI{0.66}{\meter}$ such that the circular solution does not exists if either~$\lambda \le \lambda_\mr{cr}$ or~$w\le w_\mr{cr}$.

\begin{rmk}
Note that, chosen a geometric parameter, its possible critical value does not depend on the rear speed~$v_c$. Indeed, as long as the circular motion does not exist, the only solution to system~\eqref{eq:alg_system} is the rectilinear one. In particular, the derivatives of the potential energy are always zero and we can simplify the system, which becomes independent of~$v_c$.
\end{rmk}

Therefore, in the phase space, the fixed points corresponding to the circular motion move depending on the geometric parameters, and it may happen that they coincide with the fixed point in the origin representing the rectilinear motion. Now, we want to check the stability of the rectilinear solution in this situation, that is, when the circular motion does not exist. For instance, if $\lambda=0$, we numerically see that the rectilinear solution is unstable. Hence, we would like to understand if there is a relation between the existence of circular motions and the stability of the rectilinear solution.

Let us consider the critical velocity $v_\mr{cr}$ which makes the rectilinear solution asymptotically stable. We compute its value for different choice of the geometric parameters. In Table~\ref{tab:v_critical} the dependence of~$v_\mr{cr}$ on the caster angle and the wheelbase, varied separately. We observe that, if the circular motion does not exist, then we cannot define the critical speed because the rectilinear solution appears to be unstable for any value of the rear speed.

\begin{table}[tb]
\caption{Dependence of the critical speed $v_\mr{cr}$ on the caster angle $\lambda$ and the wheelbase $w$. If the rectilinear solution is unstable, no value for $v_\mr{cr}$ is reported. The critical speed decreases monotonically when the caster angle becomes closer to the critical value $\lambda_\mr{cr}$ which determines the existence of circular solutions, while it increases monotonically if $w$ decreases to the critical value. We also note that the existence of circular motion is not sufficient for the asymptotic stability of the rectilinear solution, as there exists a value of the wheelbase $\ol w > w_\mr{cr}$ such that the rectilinear solution is unstable again.}
\label{tab:v_critical}
\centering
    \begin{tabular}{cc}
    \toprule
    \textbf{$\lambda$} & \textbf{$v_\mr{cr}(\lambda)$} \\
    \midrule
    \SI{0.2}{\degree} & - \\
    \midrule
    \SI{0.3}{\degree} & \SI{1.25}{\meter\per\second} \\
    \midrule
    \SI{0.5}{\degree} & \SI{1.35}{\meter\per\second} \\
    \midrule
    \SI{5}{\degree} & \SI{2.85}{\meter\per\second} \\
    \midrule
    \SI{10}{\degree} & \SI{3.89}{\meter\per\second} \\
    \midrule
    \SI{30}{\degree} & \SI{6.86}{\meter\per\second} \\
    \bottomrule
    \end{tabular}
    \qquad\qquad\qquad
    \begin{tabular}{cc}
    \toprule
    \textbf{$w$} & \textbf{$v_\mr{cr}(w)$} \\
    \midrule
    \SI{0.6}{\meter} & - \\
    \midrule
    \SI{0.7}{\meter} & \SI{4.18}{\meter\per\second} \\
    \midrule
    \SI{0.8}{\meter} & \SI{3.93}{\meter\per\second} \\
    \midrule
    \SI{0.9}{\meter} & \SI{3.51}{\meter\per\second} \\
    \midrule
    \SI{1.0}{\meter} & \SI{2.85}{\meter\per\second} \\
    \midrule
    \SI{1.1}{\meter} & - \\
    \bottomrule
    \end{tabular}
\end{table}



\begin{rmk}
Due to the nonlinearity of the system, the dependence of~$v_\mr{cr}$ on different geometric parameters varies. In particular, we notice that~$v_\mr{cr}(\lambda)$ decreases when the caster angle tends to~$\lambda_\mr{cr}$, that is, the rectilinear solution becomes asymptotically stable for smaller speed values. On the other hand,~$v_\mr{cr}(w)$ increases when $w\to w_\mr{cr}$, as if the system becomes \emph{less} stable in the sense that we need a higher rear speed value to have asymptotic stability. Further, we also note that the rectilinear solution is unstable again when the wheelbase is larger than $\ol w \simeq \SI{1.1}{\meter}$.
\end{rmk}

The numerics above suggest that there should be a relation between the existence of circular motions and the stability of the rectilinear motion. 
In particular, simulations hint that if we have only one fixed point in the phase space, that is, the one corresponding to the rectilinear motion, then it is unstable. 
We summarise all these observations as follows.

\begin{prb}
Is the existence of circular motions for the TMS \emph{necessary} condition for the asymptotic stability of the rectilinear motion? That is, if there exists no circular solution, is the rectilinear motion unstable for any value of the rear speed?
\end{prb}



\section{Numerical simulations}


Numerical solutions of the nonlinear TMS show different and interesting dynamics which depend on the initial conditions and the geometric parameters chosen. Also, we are interested in comparing our results with those obtained in~\cite{meijaard:science} for the theoretical and experimental TMS. On the one hand, we want to check whether the linearised theoretical model exhibits dynamical behaviours predicted also by our nonlinear version. On the other hand, we want to see if the nonlinear model derived in this paper can accurately describe the behaviour of the real system. 

\subsection{Theoretical TMS}

We present two different simulations. In the first we choose an initial rear speed such that the rectilinear motion is asymptotically stable, while the second simulation will show a different kind of stability. The geometric parameters for the theoretical TMS are listed in Table~\vref{tab:values1}.

\subsubsection{Asymptotically stable rectilinear motion}

\begin{figure}[htb]
    \centering
    \begin{subfigure}[b]{0.4\textwidth}
        \includegraphics[width=\textwidth]{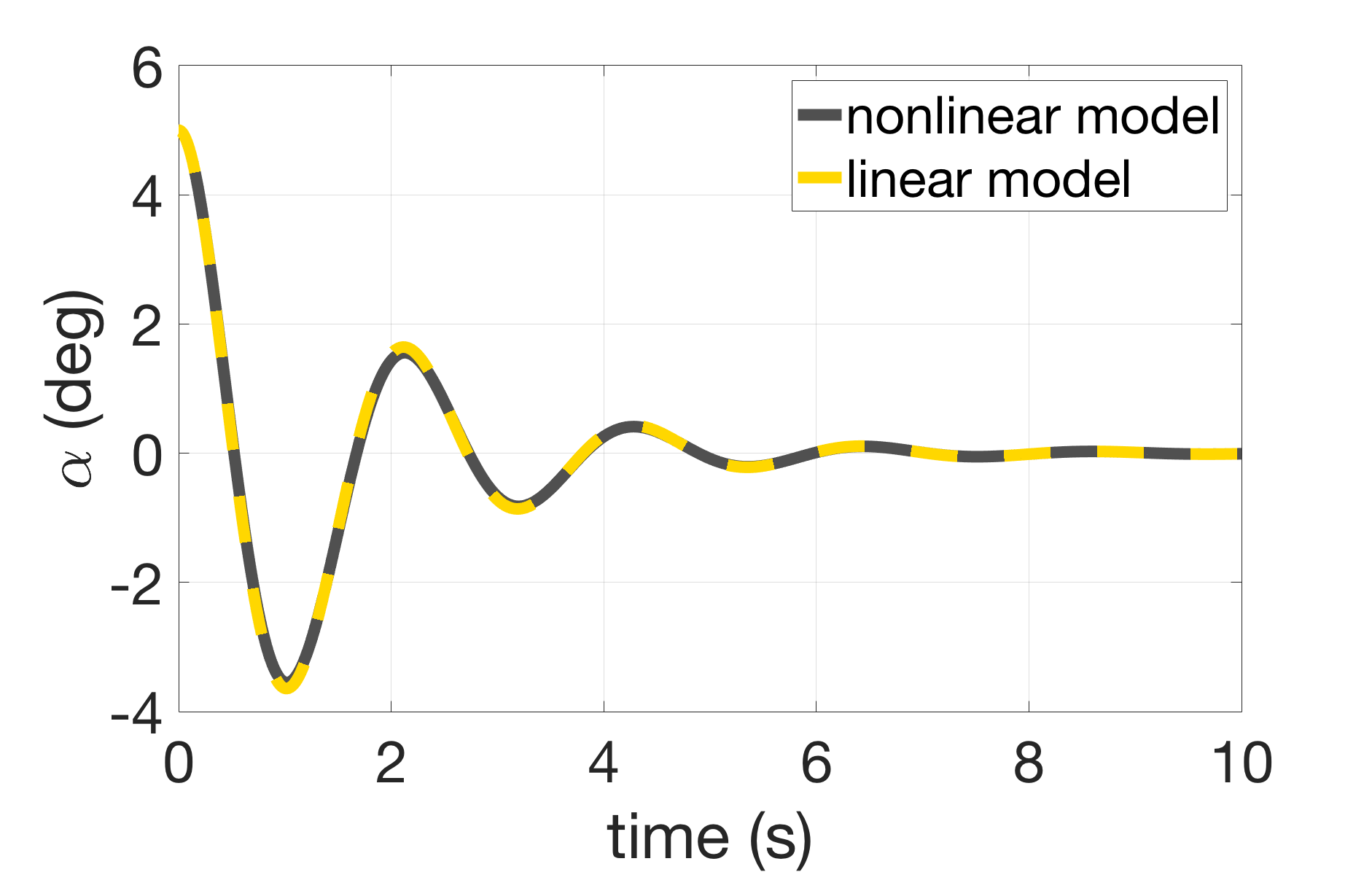}
        \caption{Evolution of the roll angle.}
    \end{subfigure}
    ~
    \begin{subfigure}[b]{0.4\textwidth}
        \includegraphics[width=\textwidth]{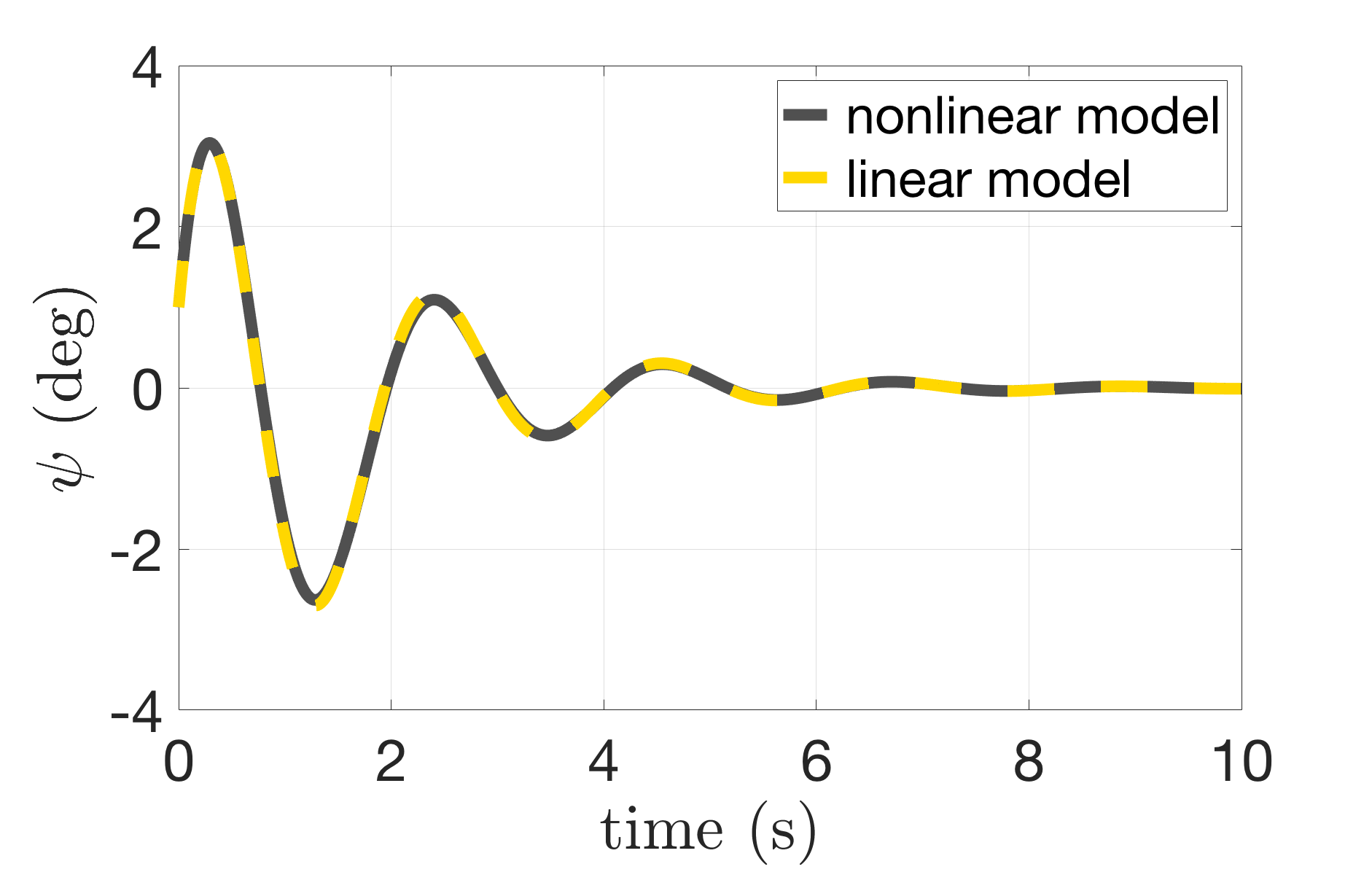}
        \caption{Evolution of the steering angle.}
    \end{subfigure}
    
    \begin{subfigure}[b]{0.4\textwidth}
        \includegraphics[width=\textwidth]{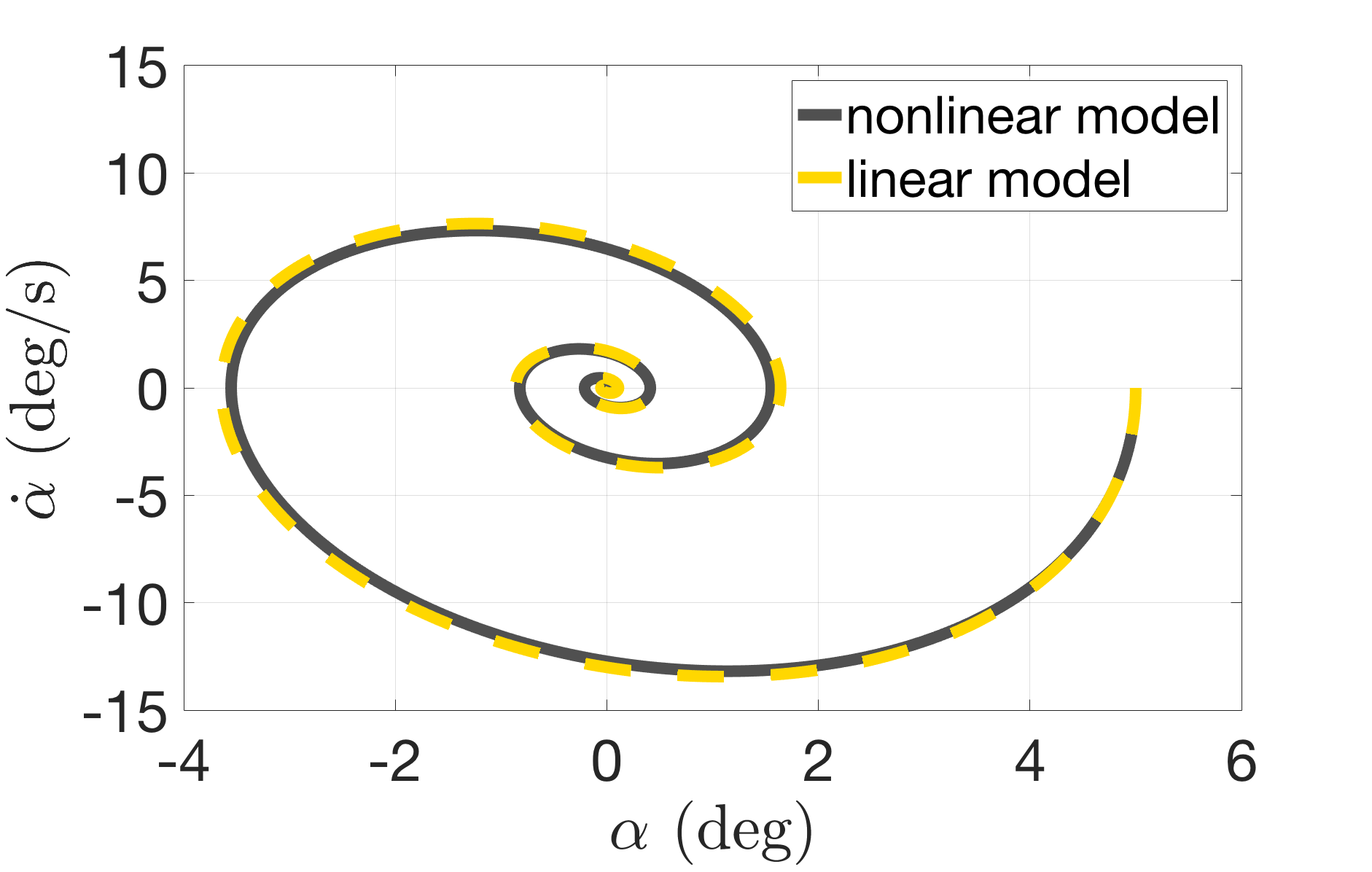}
        \caption{Phase plane for the roll angle.}
        \label{fig:phase_alpha_test1}
    \end{subfigure}
    ~
    \begin{subfigure}[b]{0.4\textwidth}
        \includegraphics[width=\textwidth]{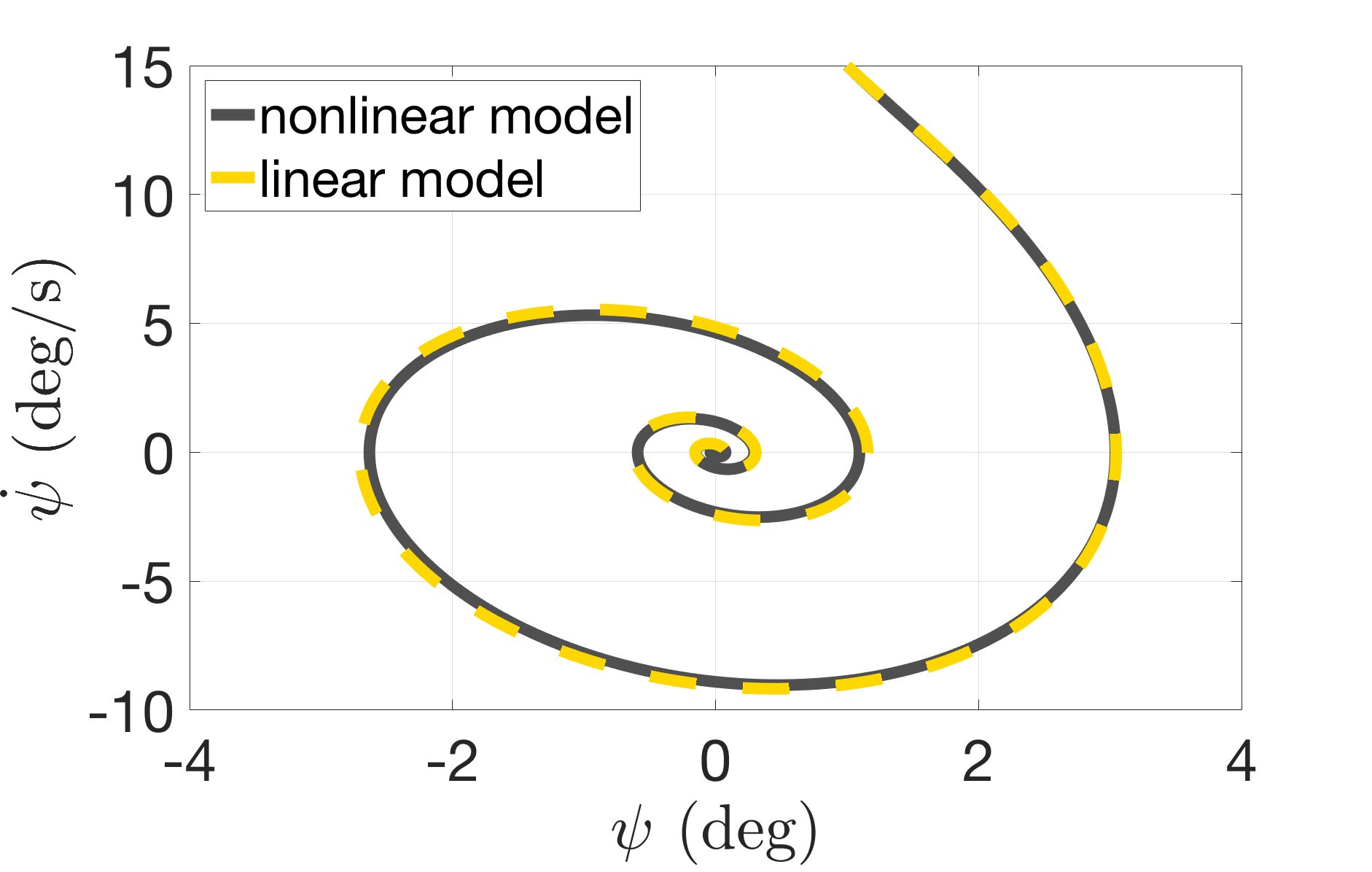}
        \caption{Phase plane for the steering angle.}
        \label{fig:phase_psi_test1}
    \end{subfigure}
    
     \begin{subfigure}[b]{0.4\textwidth}
        \includegraphics[width=\textwidth]{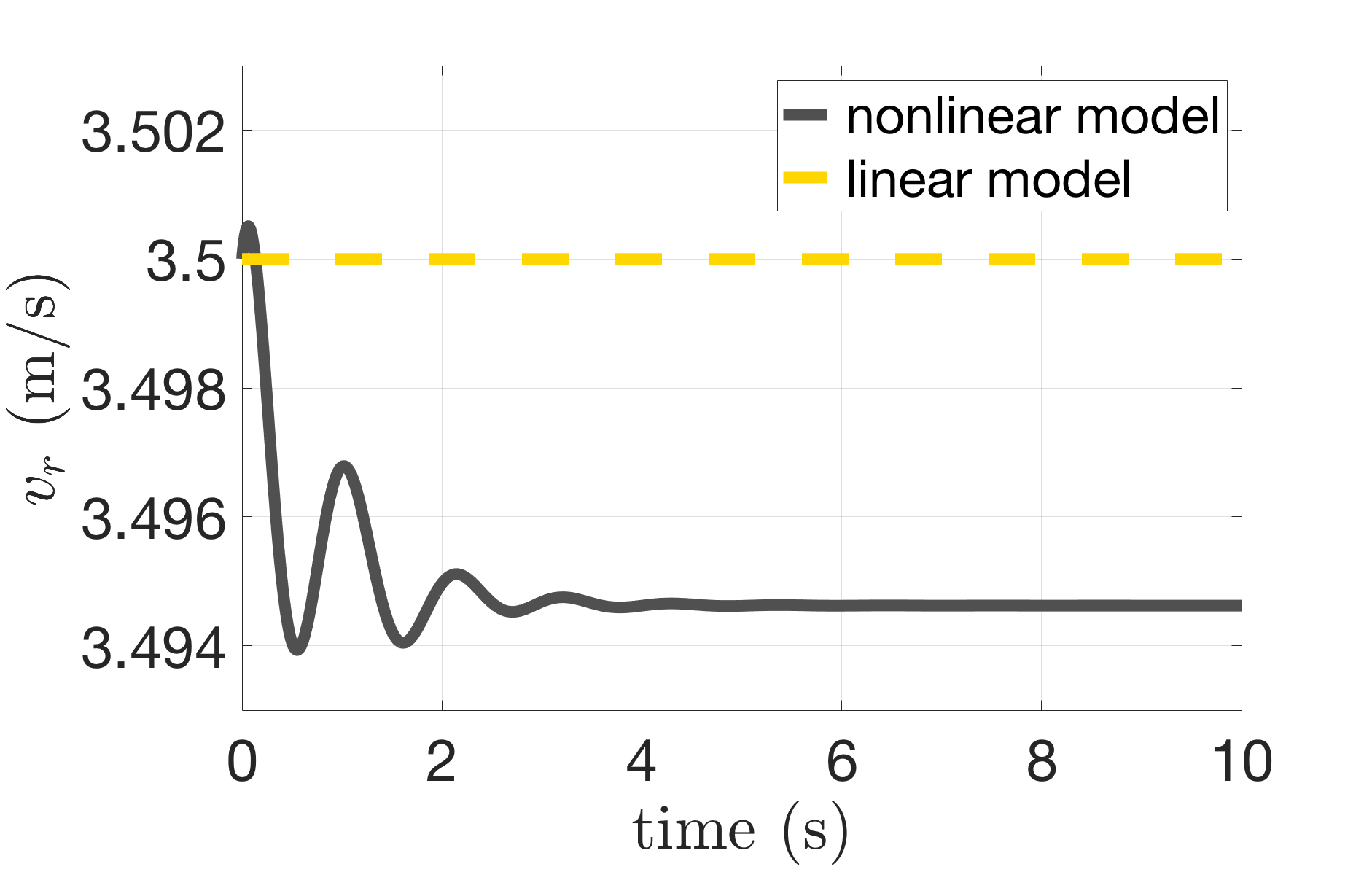}
        \caption{Speed of the rear contact point.}
        \label{fig:vr_test1}
    \end{subfigure}
    ~
    \begin{subfigure}[b]{0.4\textwidth}
        \includegraphics[width=\textwidth]{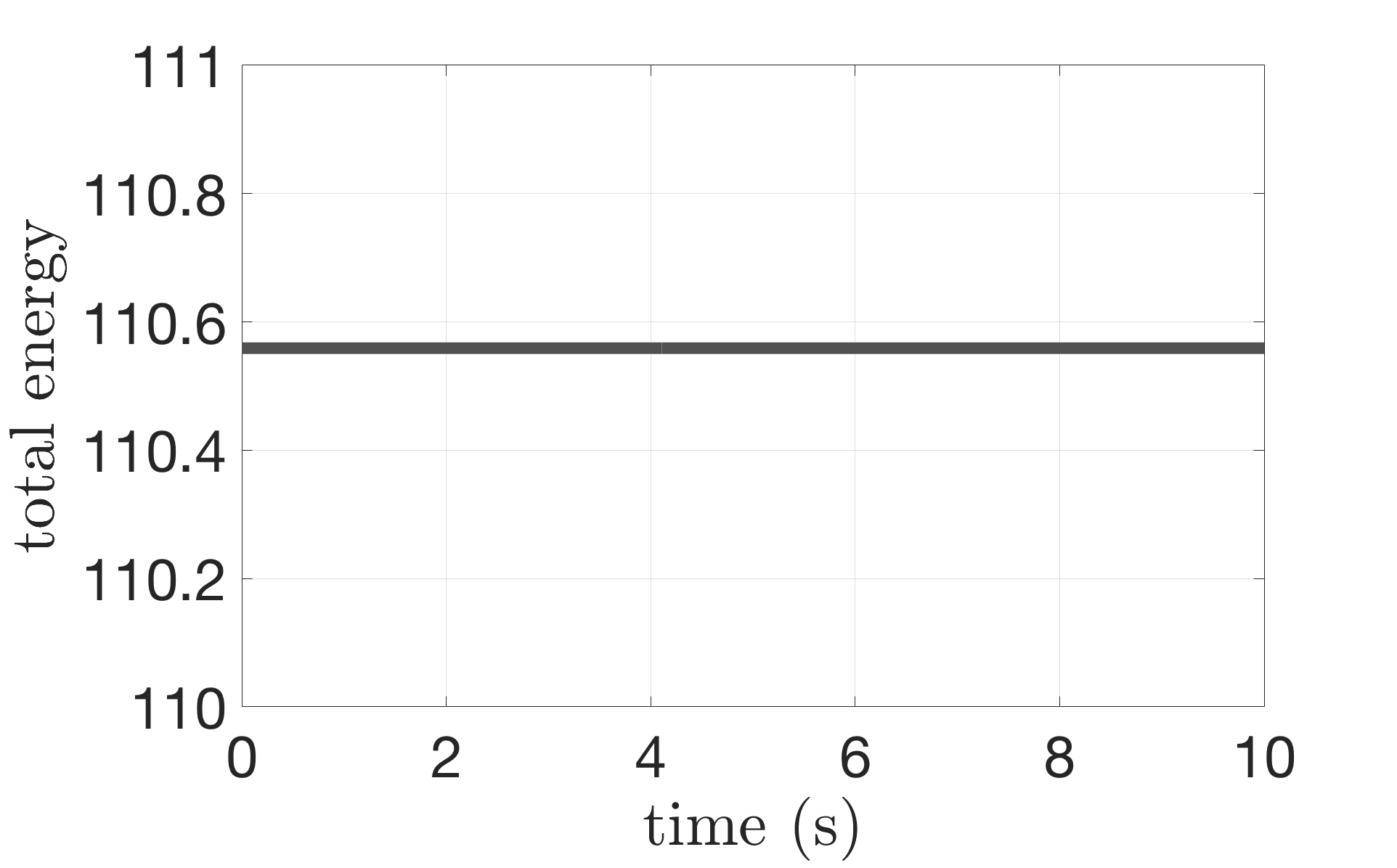}
        \caption{Total energy of nonlinear model.}
        \label{fig:energy_test1}
    \end{subfigure}
    \caption{Dynamical behaviour of the linear TMS (dashed line) and the nonlinear one (solid line) for a $v_r(0)>v_\mr{cr}$. The system presents stability according to both the models and tends to the rectilinear motion after a transient time.}
    \label{fig:test1}
\end{figure}

Given the geometric parameters of Table~\ref{tab:values1}, we know from the previous Section that there exists a critical speed $v_\mr{cr}$ such that the rectilinear motion is asymptotically stable. Hence, we choose~$v_r(0)>v_\mr{cr}$ and the initial roll and steering angles in the \emph{basin of attraction} of the rectilinear motion, which is the set of points such that initial conditions chosen in this set dynamically evolve to the rectilinear motion. 
For instance, by choosing
\begin{align*}
&\begin{system}
\alpha(0) = \SI{5}{\degree}, \\
\psi(0) = \SI{1}{\degree}, \\
s_r(0) = \SI{0}{\metre},
\end{system}
&
&\begin{system}
\dot\alpha(0) = \SI{0}{\degree\per\second}, \\
\dot\psi(0) = \SI{15}{\degree\per\second}, \\
v_r(0) = \SI{3.5}{\metre\per\second}.
\end{system}
\end{align*}
the numerical solution is asymptotically stable. With these initial conditions, the same behaviour is shown also by the linear model from~\cite{meijaard:science}. In Figure~\ref{fig:test1} we plot the evolution of the roll angle and the steering angle given by the two models, together with their phase portraits.

In the phase portraits~\ref{fig:phase_alpha_test1} and~\ref{fig:phase_psi_test1} we observe an \emph{ingoing spiral} for both the models, that is, we have a spiral sink in which the oscillations decay. This behaviour is related to the presence of the pair of complex eigenvalues of the Jacobian matrix which have nonzero imaginary part, as computed before. The difference between the two model appears in the transient time, and it becomes more visible when the initial conditions for the angular coordinates are not close to those characteristic of the rectilinear motion.

Moreover, running different simulations with the nonlinear model, we notice that its basin of attraction for the rectilinear solution appears to change with the geometric parameters of the bicycle. Therefore, we formulate the following question.

\begin{prb}
Can we determine the basin of attraction of the nonlinear model for the rectilinear motion? How does it depend on the geometric parameters of the model?
\end{prb}



Finally, we note that in the simulation the total energy of the system is conserved, as shown in Figure~\ref{fig:energy_test1}. This is in compliance with the theoretical aspect that systems with nonholonomic constraints linear in the velocities conserve the total energy, see~\cite{bloch:nonholo}.

\subsubsection{Existence of a limit cycle}

\begin{figure}[htb]
    \centering
    \begin{subfigure}[b]{0.4\textwidth}
        \includegraphics[width=\textwidth]{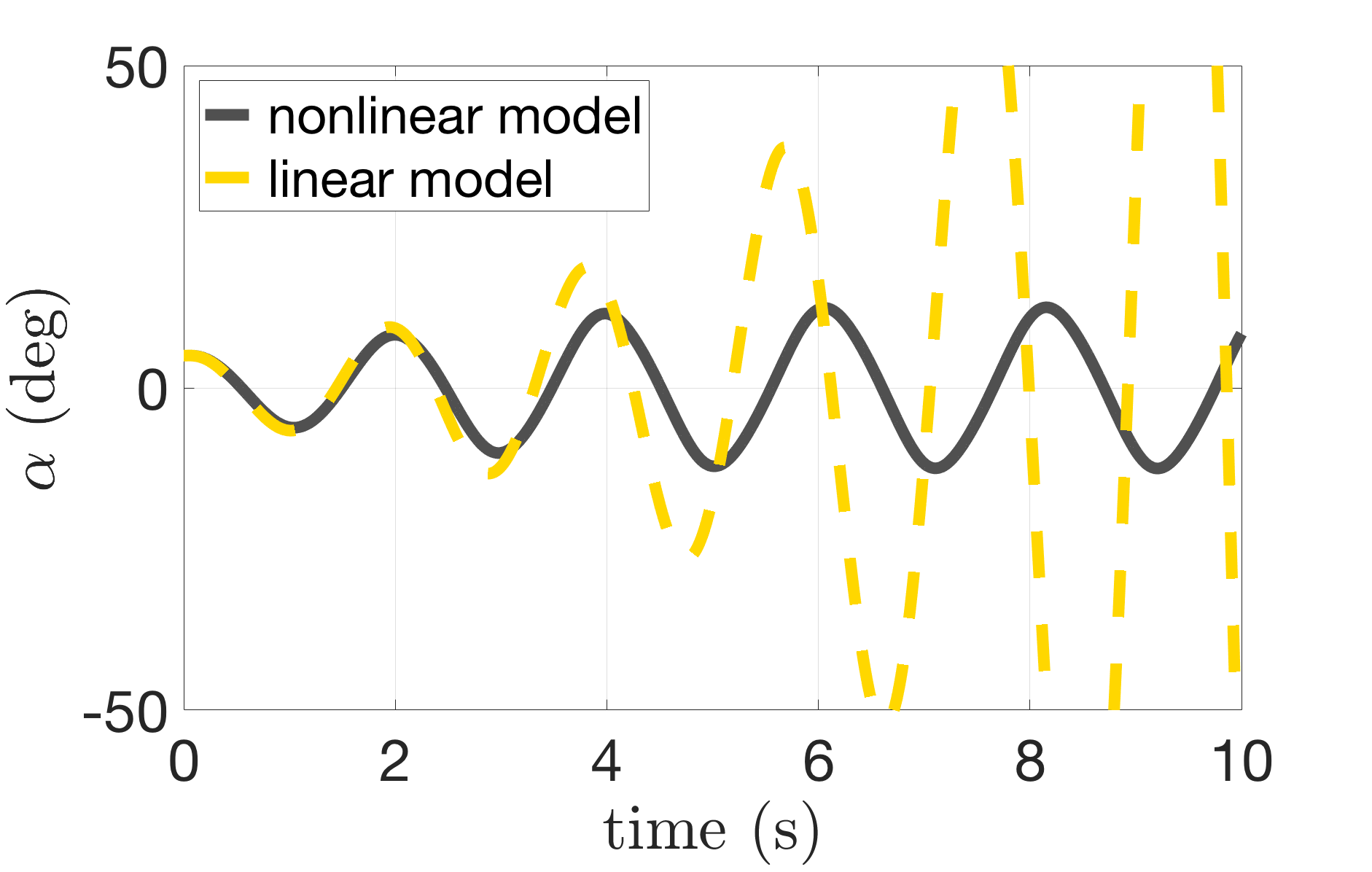}
        \caption{Evolution of the roll angle.}
    \end{subfigure}
    ~
    \begin{subfigure}[b]{0.4\textwidth}
        \includegraphics[width=\textwidth]{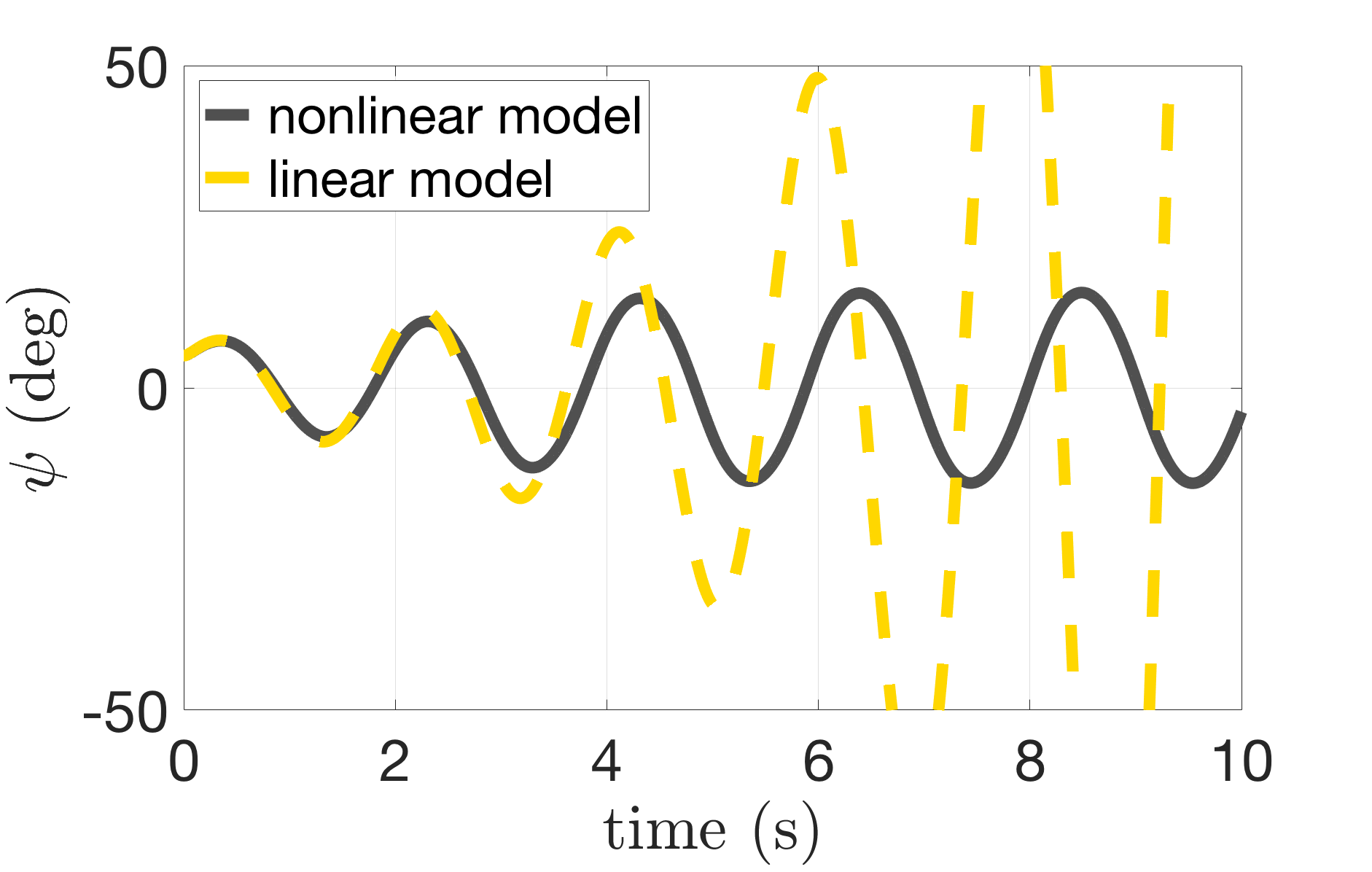}
        \caption{Evolution of the steering angle.}
    \end{subfigure}
    
    \begin{subfigure}[b]{0.4\textwidth}
        \includegraphics[width=\textwidth]{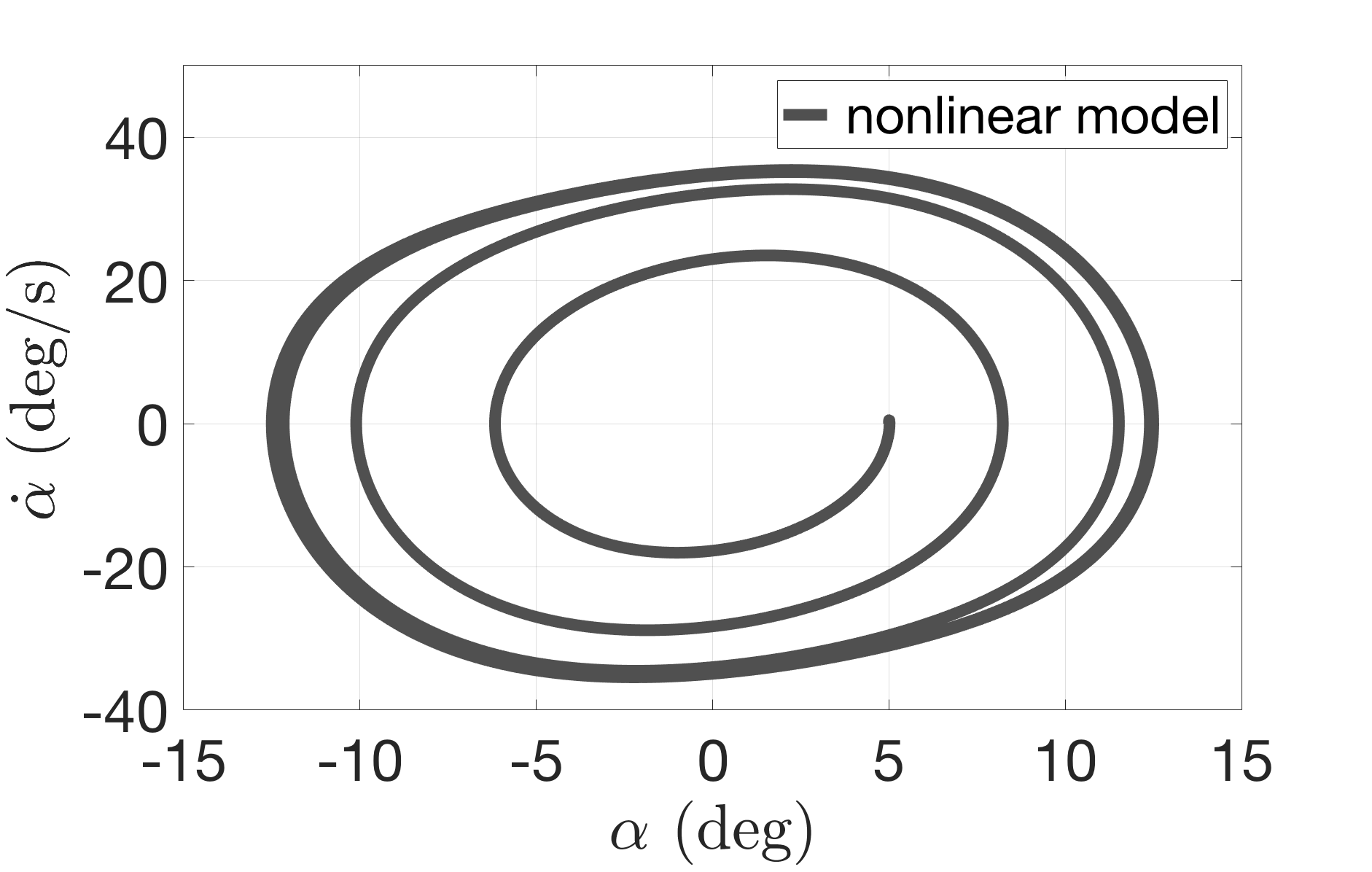}
        \caption{Phase plane for the roll angle.}\label{fig:phase_alpha_test2}
    \end{subfigure}
    ~
    \begin{subfigure}[b]{0.4\textwidth}
        \includegraphics[width=\textwidth]{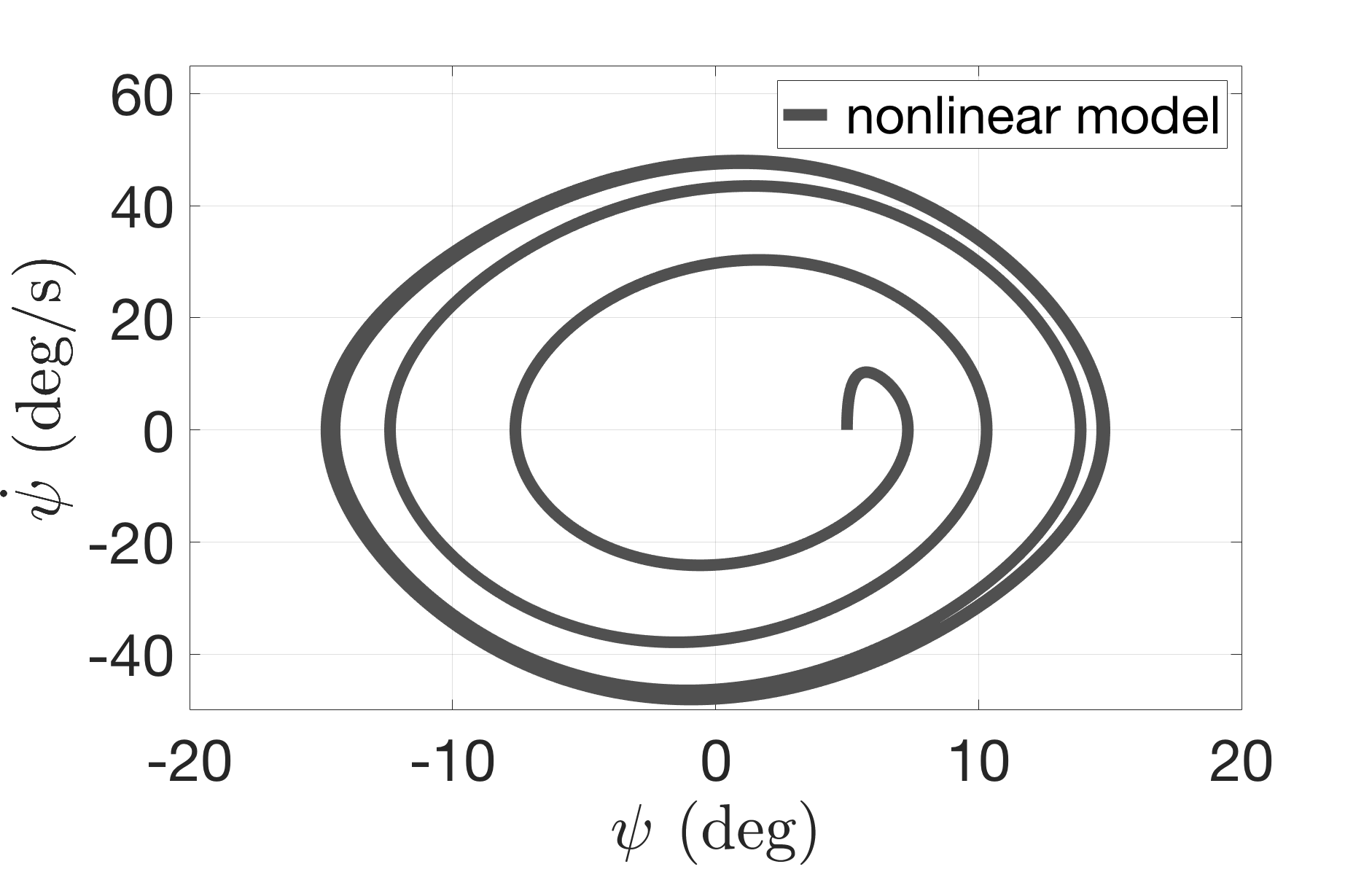}
        \caption{Phase plane for the steering angle.}\label{fig:phase_psi_test2}
    \end{subfigure}
    
     \begin{subfigure}[b]{0.4\textwidth}
        \includegraphics[width=\textwidth]{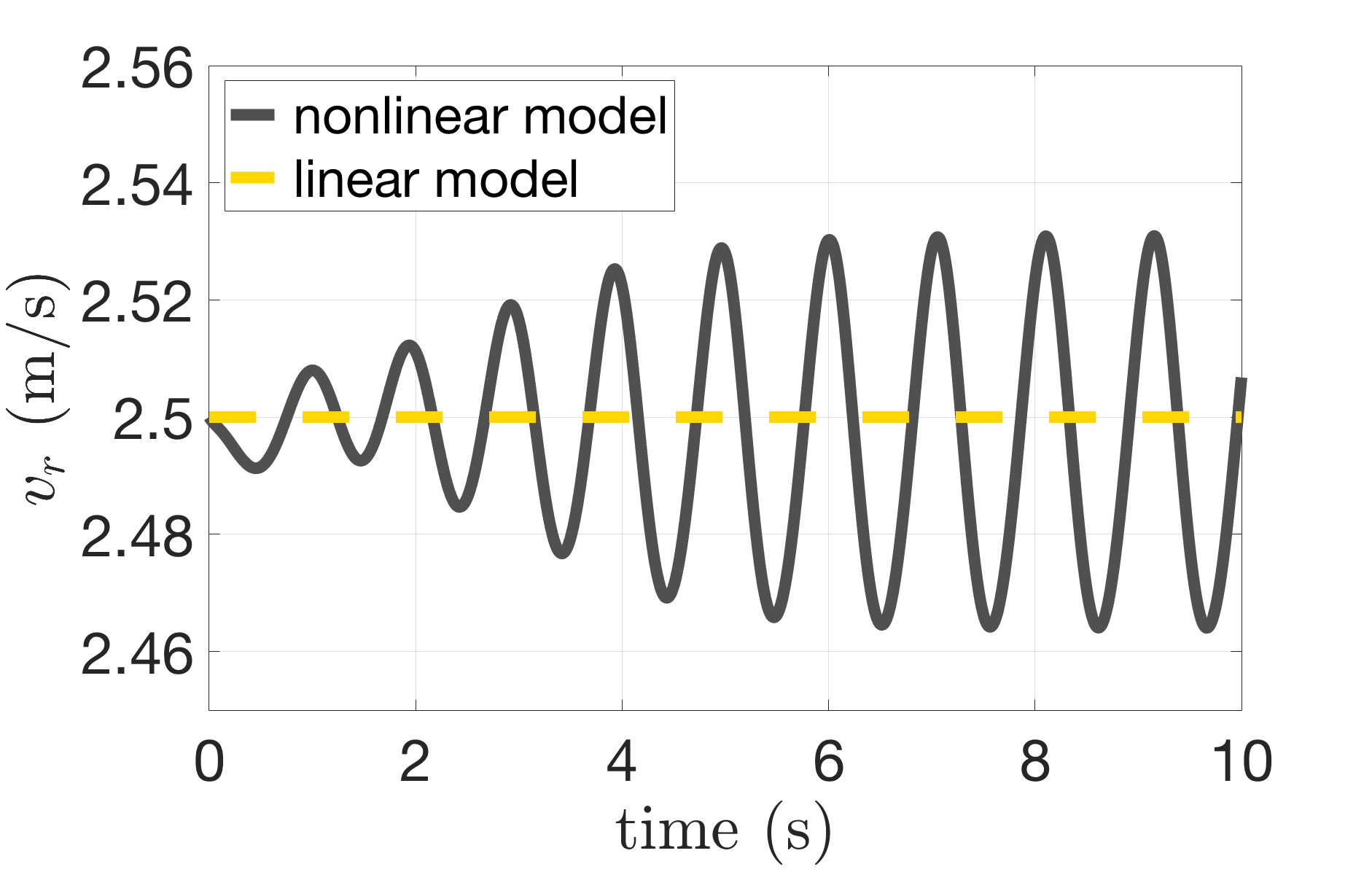}
        \caption{Speed of the rear contact point.}
        \label{fig:vr_test2}
    \end{subfigure}
    ~
    \begin{subfigure}[b]{0.4\textwidth}
        \includegraphics[width=\textwidth]{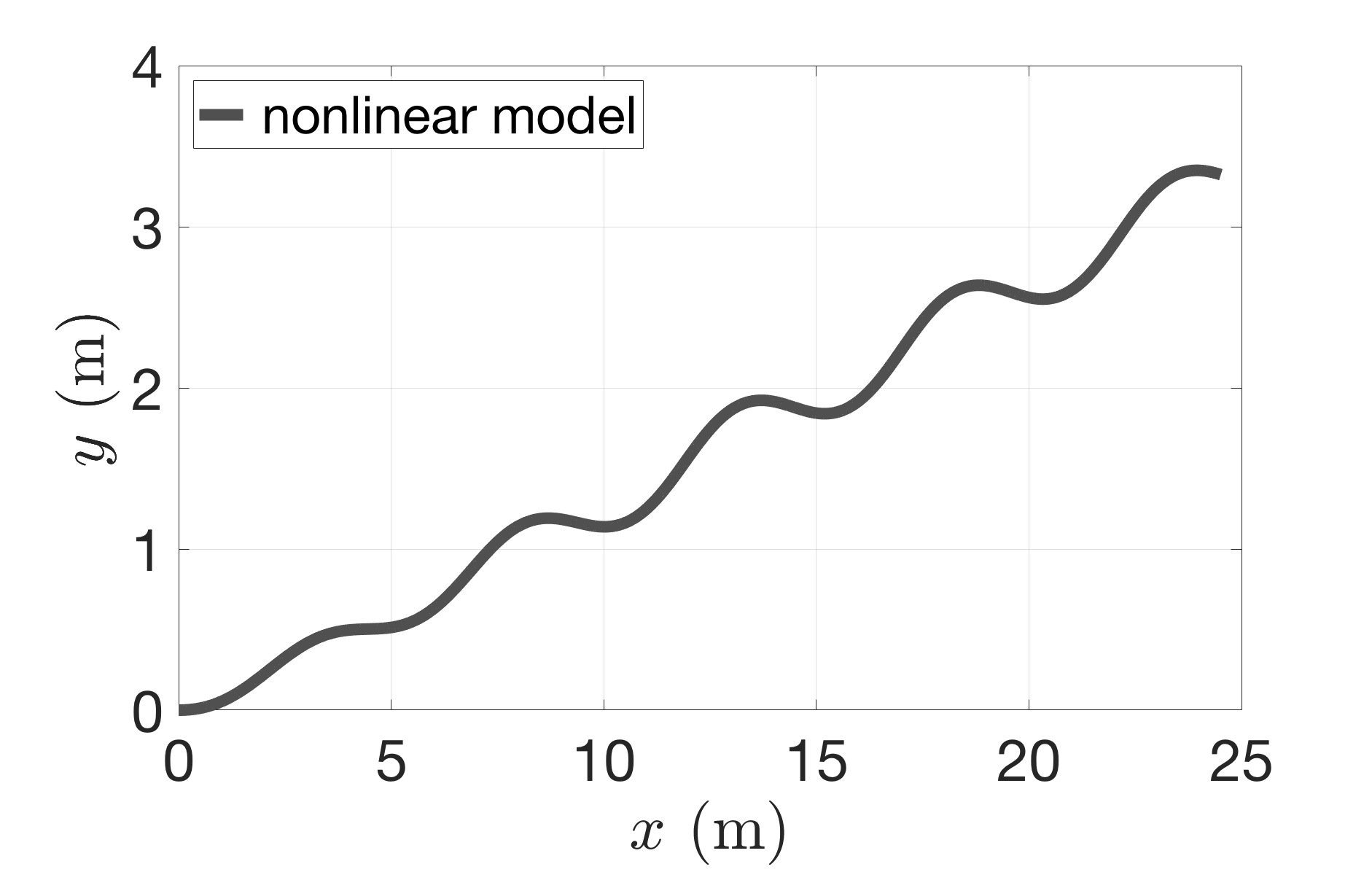}
        \caption{Planar path for the nonlinear model.}
        \label{fig:xy_test2}
    \end{subfigure}
    \caption{Dynamical behaviour of the linear TMS (dashed line) and the nonlinear one (solid line) when the nonlinear system has a limit cycle. Unlike our nonlinear model, the linear one predicts an unstable behaviour.}
    \label{fig:test2}
\end{figure}

In this second simulation we want to show what happens if the rectilinear solution is not asymptotically stable, that is, if the initial rear speed is lower than the critical value~$v_\mr{cr}$. In particular, we choose the following initial conditions
\[
\begin{system}
\alpha(0) = \SI{5}{\degree}, \\
\psi(0) = \SI{5}{\degree}, \\
v_r(0) = \SI{2.5}{\metre\per\second},
\end{system}
\]
where the remaining values are zero.
In Figure~\ref{fig:test2} we plot the evolution of the roll and steering angle as well as their phase portrait. We note that the difference between the dynamics of the linear and the nonlinear model is evident. The nonlinear model presents a \emph{limit cycle}, that is, it exhibits a stable behaviour. This is a behaviour typical of a nonlinear system, which cannot be shown by a linear model. Indeed, the linear model is unstable, and we can check that the Jacobian matrix evaluated along the rectilinear motion has now a pair of complex eigenvalues with positive real part, that is, we have a  \emph{spiral source} corresponding to growing oscillation of the system.



Figure~\ref{fig:phase_alpha_test2} and~\ref{fig:phase_psi_test2} show the projection of the trajectory on the phase planes for the roll and the steering angle, respectively. We have not plotted the phase portraits of the linear model as it is not interesting. In the two pictures we can clearly observe how the solution of the nonlinear system spirals into the limit cycle as time approaches infinity, and that it is \emph{symmetric} with respect to the origin due to the symmetry of the system. However, the limit cycle appears to be \emph{semi-stable}, that is, we can find neighbouring trajectories which spiral into the limit cycle as time approaches negative infinity. 

Even in this case, given an initial speed for which the limit cycle exists, we want to determine what is the basin of attraction of the limit cycle and how it is related to the geometric parameters of the bicycle. From simulations, it seems that the answer may be related to the circular motion of the TMS.

If we consider the four dimensional phase space parametrised by $\xi=(\alpha,\psi, \dot\alpha,\dot\psi)$, let $\alpha_l$ and~$\psi_l$ be the angles in the limit cycle corresponding to $\dot\alpha_l=0$ and~$\dot\psi_l=0$, respectively. Moreover, we compute $\alpha_c$ and $\psi_c$ as solutions of system~\eqref{eq:alg_system} taking the rear speed given by the limit cycle when $\alpha=\alpha_l$ and $\psi=\psi_l$, respectively. Then, the numerics show that the limit cycle exists and it is semi-stable if either~$\abs{\alpha_l}<\abs{\alpha_c}$ or~$\abs{\psi_l}<\abs{\psi_c}$. Again, due to the symmetry of the system, it does not make any difference if we take positive or negative angle, as long as the choice is consistent. The problem is summarised in the following question.

\begin{prb}
Can we determine the basin of attraction of the limit cycle? How does it depend on the geometric parameters of the bicycle and, possibly, on the circular motion?
\end{prb}

We finally observe that the frequency of the rear speed oscillations is practically double the frequencies of either the roll or steering angles, as shown in Figure~\ref{fig:vr_test2}, and possibly this pattern is typical of the limit cycle. about the evolution of the rear speed.

\subsubsection{Additional remarks}

From the two different simulations reported above we can figure out the following behaviours exhibited by our nonlinear model. Given the geometric parameters in Table~\ref{tab:values1}, we determine a critical value $v_\mr{cr}$ such that the rectilinear motion, seen as a stable fixed point in the four-dimensional phase space, is stable for any initial rear speed~$v_r(0)> v_\mr{cr}$. Then, we observed that the system has a semi-stable limit cycle for $v_\mr{lc}<v_r(0)<v_\mr{cr}$, where $v_\mr{lc}$ is the lower critical speed such that the limit cycle exists. Finally, still decreasing the initial value of the rear speed, the limit cycle disappears and the system is unstable. This phenomenology reminds a \emph{Hopf bifurcation} of the system. We state this observations in the following question.

%

\begin{prb}
Given a set of geometric parameters for which 
the rectilinear solution is eventually stable, does the system undergo a Hopf bifurcation with respect to the rear speed $v_r(0)$? Being $v_\mr{cr}$ and $v_\mr{lc}$ the upper and lower bound for the existence of the limit cycle, how these values depend on the geometric parameters of the TMS?
\end{prb}

\subsection{Experimental TMS}

We now consider the experimental TMS whose construction is described in~\cite{meijaard:science}. The main feature of this bicycle is that the wheels are designed in order to reduce gyroscopic effects almost to zero, since the main assumption of the theoretical model was to consider the wheels as point masses. However, the experimental TMS presents distributed masses for the rear and front frames, which were not included in the original theoretical model introduced in~\cite{meijaard:science}.

Nevertheless, the authors showed that the linear theoretical model approximates well the behaviour of the experimental TMS when the rectilinear solution is asymptotically stable and the initial conditions are chosen in its basin of attraction. This is a consequence of the asymptotic stability of the rectilinear solution. The difference between the linear model and the real dynamics in the transient period before the rectilinear motion is reported by the authors as a \SI{20}{\hertz} shimmy motion in the steering assembly that is not predicted by the low-dimensional linearised model.

\begin{table}[tb]
\caption{Nonzero geometric parameters of the experimental TMS as in~\cite{meijaard:science}. The coordinates of the two centres of mass are expressed in the trivial configuration with respect to the fixed reference frame $\Sigma$ with $O\equiv A$. We chose $Z$ to be directed upward, thus $z_2$ and $z_3$ have signs opposite to those in the original paper. Also, the moments of inertia are expressed in our local reference systems, the height of the centres of mass is reduced by the radius of the wheels, and the trail is not listed since it has not been introduced in our model.}
\label{tab:values2}
\centering
    \begin{tabular}{lll}
    \toprule
    \textbf{Symbol} & \textbf{Parameter} & \textbf{Value} \\
    \midrule
    $w$ & wheelbase & \SI{0.750}{m} \\
    \midrule
    $\lambda$ & caster angle & \SI{7}{\degree} \\
    \midrule
    $m_1$ & rear wheel mass & \SI{0.342}{\kilo\gram} \\
    \midrule
    $(x_G, z_G)$ & \specialcell[c]{rear frame\\centre of mass} & (\SI{0.5044}{m}, \SI{0.4279}{m}) \\
    \midrule
    $m_2$ & rear frame mass & \SI{6.425}{\kilo\gram} \\
    \midrule
    $\begin{pmatrix}
    I_{xx2}  & 0      & I_{xz2} \\
    0       & I_{yy2} & 0       \\
    I_{xz2} & 0      & I_{zz2}  \\
    \end{pmatrix}$
    & \specialcell[c]{rear frame\\inertia tensor} &
    $\begin{pmatrix}
    0.06460 & 0      & 0.23102 \\
    0       & 2.59262 & 0       \\
    0.23102 & 0      & 2.54642  \\
    \end{pmatrix}$
    \si{\kilo\gram\metre^2} \\
    \midrule
    $(x_H, z_H)$ & \specialcell[c]{front frame\\centre of mass} & (\SI{0.7338}{m}, \SI{0.3022}{m}) \\
    \midrule
    $m_3$ & front frame mass & \SI{2.412}{\kilo\gram} \\
    \midrule
    $\begin{pmatrix}
    I_{xx3}  & 0      & I_{xz3} \\
    0       & I_{yy3} & 0       \\
    I_{xz3} & 0      & I_{zz3}  \\
    \end{pmatrix}$
    & \specialcell[c]{front frame\\inertia tensor} &
    $\begin{pmatrix}
    0.03797  & 0       & -0.00393 \\
    0        & 0.03807 & 0        \\
    -0.00393 & 0       & 0.00185  \\
    \end{pmatrix}$
    \si{\kilo\gram\metre^2} \\
    \midrule
    $m_4$ & front wheel mass & \SI{0.342}{\kilo\gram} \\
    \bottomrule
    \end{tabular}
\end{table}

Since our nonlinear model includes also distributed masses, we now want to compare its dynamics with the experimental results presented in~\cite{meijaard:science}. 
In the following we present the same analysis we did previously for the theoretical TMS, that is, checking the stability of the rectilinear solution for the experimental TMS and running few simulations to compare the predicted dynamics with the real one. In this case, the nonzero geometric parameters are those listed in Table~\ref{tab:values2}, and again we use relations~\eqref{eq:trig_rel}.

Note that we need to express the inertia tensor with respect to our reference system. Indeed, in~\cite{meijaard:science} the moments of inertia are computed with respect to a different local reference frame, and the axes rotation is given by the matrices
\begin{align*}
R_2(\mu) &=
\begin{pmatrix}
\cos \mu & 0 & -\sin\mu \\
0 & -1 & 0 \\
-\sin\mu & 0 & -\cos\mu
\end{pmatrix}
& &\text{and} &
R_3(\lambda) &=
\begin{pmatrix}
\cos \lambda & 0 & -\sin\lambda \\
0 & -1 & 0 \\
-\sin\lambda & 0 & -\cos\lambda
\end{pmatrix}
\end{align*}
for the rear and front frame, respectively. The moments of inertia are then obtained through the transformation $\sigma_i = R_i^T \sigma_i^K R_i$, $i=2, 3$, where $\sigma_i^K$ is the inertia tensor as in~\cite{meijaard:science}. The values in Table~\ref{tab:values2} are expressed in our reference system.

\subsubsection{Stability analysis of the rectilinear motion}

\begin{figure}[htb]
\centering
\includegraphics[width=0.65\textwidth]{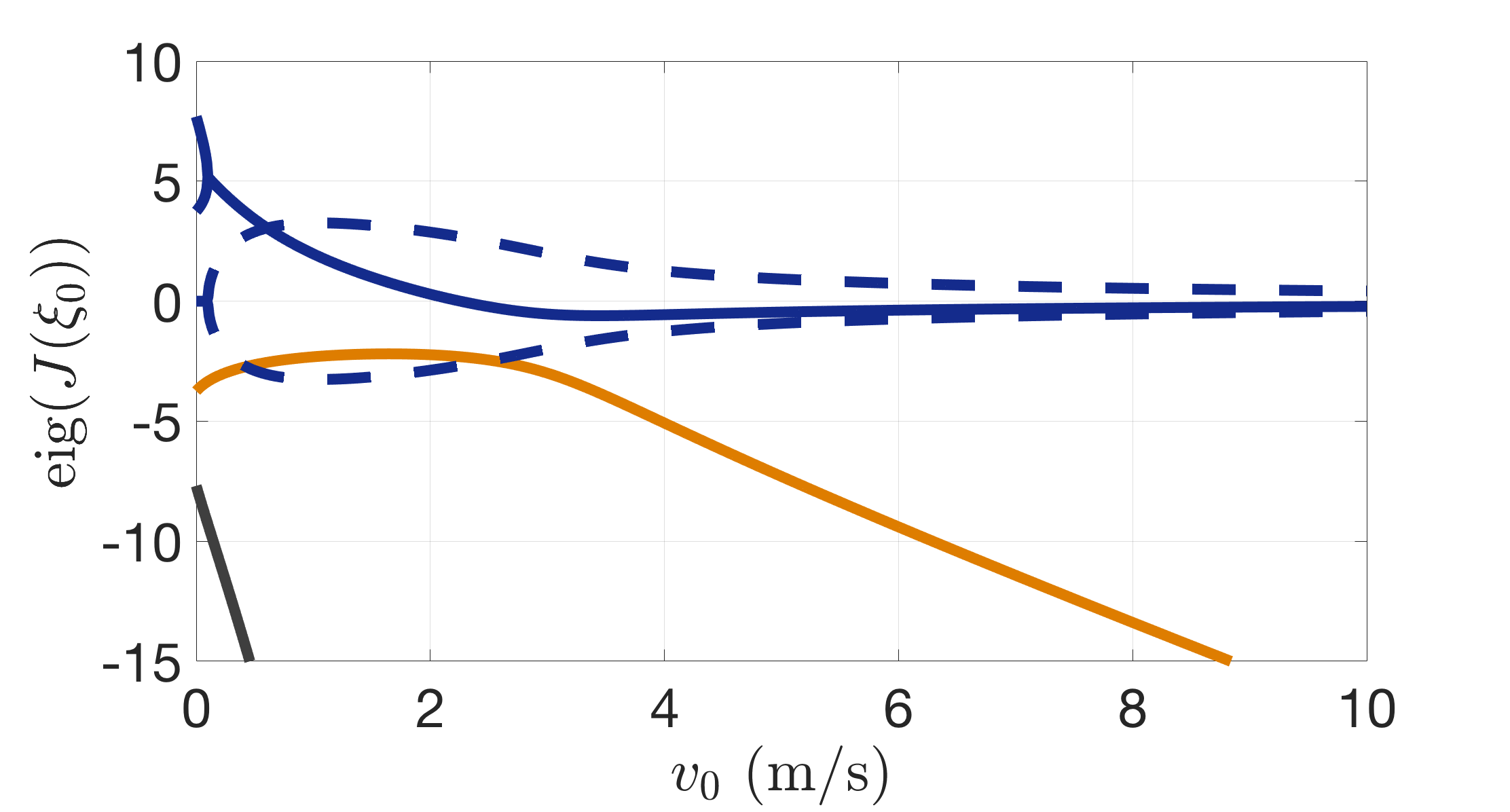}
\caption{Dependence of the four eigenvalues of the Jacobian $J(\xi_0)$ on the rear speed $v_0$ for the experimental TMS. Real (solid line) and imaginary (dashed line) part of the eigenvalues is plotted. As long as~$v_0< \SI{0.1}{\meter\per\second}$, all the eigenvalues are real, while for $v_0\ge \SI{0.1}{\meter\per\second}$ we have two real eigenvalues and a pair of complex conjugate eigenvalues. The real eigenvalues are negative for every value of the rear speed, while the complex conjugate pair has negative real part only if~$v_0\ge\SI{2.26}{\meter\per\second}$. }
\label{fig:eigen_rect_expTMS}
\end{figure}

We analyse the stability of the rectilinear motion following the procedure explained before. In particular, we linearise the equations of motions \eqref{eq:alpha_expl}, \eqref{eq:psi_expl} and~\eqref{eq:vr_expl} about the rectilinear solution $q_0(t)$, and we consider only the angular coordinates. In Figure~\ref{fig:eigen_rect_expTMS} we plot the four eigenvalues of the Jacobian matrix $J(\xi_0)$ depending on the rear speed~$v_0$. We observe that, for the experimental TMS, we initially have only real eigenvalues, two of them positive and the other two negative. Then, for $v_0\ge \SI{0.1}{\meter\per\second}$, the two positive eigenvalues becomes a pair of complex eigenvalues, and their common real part is negative only if $v_0$ is larger than the critical value $v_\mr{cr} \simeq \SI{2.26}{\meter\per\second}$, that is, the rectilinear solution becomes asymptotically stable.

Note that the critical value got for the experimental TMS is slightly smaller than the one found for the theoretical TMS, that is, for this choice of geometric parameters the rectilinear solution becomes asymptotically stable for smaller value of the rear speed. Furthermore, the critical value is the same as in~\cite{meijaard:science}, reported to be \SI{2.3}{\meter\per\second}, even if the authors used the full 25 parameters bicycle model described in~\cite{meijaard:bike} and~\cite{meijaard:linearized} for simulating the experimental TMS, while in our nonlinear model we assumed that both wheels are point masses.

\begin{rmk}
As for the theoretical TMS, also in this case we can look at the circular motions of the system. We do not report all the details, since the results are similar to those we got for the theoretical TMS. However, we remark that the geometric parameters which define the experimental TMS are such that there exists a circular solution for any value of the rear speed. This agrees with the necessity of circular motions in order to have an asymptotically stable rectilinear solution.
\end{rmk}

\subsubsection{Choice of initial conditions and simulations}

We simulate the dynamics of the experimental TMS taking the initial conditions from the experimental data sheet that can be found on the web page~\url{http://bicycle.tudelft.nl/stablebicycle/}. The measuring system realised for the experimental TMS measured the rear frame lean, that is, the roll angle $\alpha$, and the yaw angle $\theta$, together with their rates. Note that the yaw angle does not appear in the equations of motion since it is a fibre coordinate determined through the nonholonomic constraints, but we need the initial value of the steering angle and its rate. We obtained the required values from the measured data.

By using relation~\eqref{eq:front_constraints2} we can determine the steering angle directly from the data, noting that the yaw angle and steering angle in~\cite{meijaard:bike} were defined with opposite sign with respect to our definition. Then, since there is no useful relation, we use the central difference
\[
\dot\psi(t)\simeq\frac{\psi(t+h)-\psi(t-h)}{2h}
\]
to determine the initial angular speed of the steering angle. In the experiment the bicycle was hit sideways when its forward speed was $v_r(0)=\SI{3.6}{\meter\per\second}$, hence we start our simulation with the data measured after this sideways hit. According to the data, at time $t_1=\SI{0.0398}{\second}$, that is, after the hit, we have $\alpha(t_1) = \SI{0.01051}{\radian}$, $\dot\alpha(t_1) = \SI{0.41187}{\radian}$ and $\dot\theta(t_1) = \SI{0.20703}{\radian}$, and we compute $\psi(t_1)=\SI{-0.04343}{\radian}$ as initial value of the steering angle. Then, we compute the steering angle at measured time $t_0 = \SI{0}{\second}$ and $t_2= \SI{0.0798}{\second}$, and using the central difference we find $\dot\psi(t_1) = \SI{-1.01642}{\radian}$. Note that the time step is not constant, but we assume it constant as the difference is small. Therefore, shifting the time, we choose the initial conditions
\begin{align*}
&\begin{system}
\alpha(0) = \SI{0.01051}{\radian}, \\
\psi(0) = \SI{-0.04343}{\radian}, \\
s_r(0) = \SI{0}{\metre},
\end{system}
&
&\begin{system}
\dot\alpha(0) = \SI{0.41187}{\radian\per\second}, \\
\dot\psi(0) = \SI{-1.01642}{\radian\per\second}, \\
v_r(0) = \SI{3.6}{\metre\per\second}.
\end{system}
\end{align*}
Since the initial rear speed is larger than the critical value $v_\mr{cr}$, the system shows asymptotic stability. The nonzero roll and steering rates are due to the lateral perturbation applied during the experiment.

\begin{figure}[htb]
    \centering
    \begin{subfigure}[b]{0.45\textwidth}
        \includegraphics[width=\textwidth]{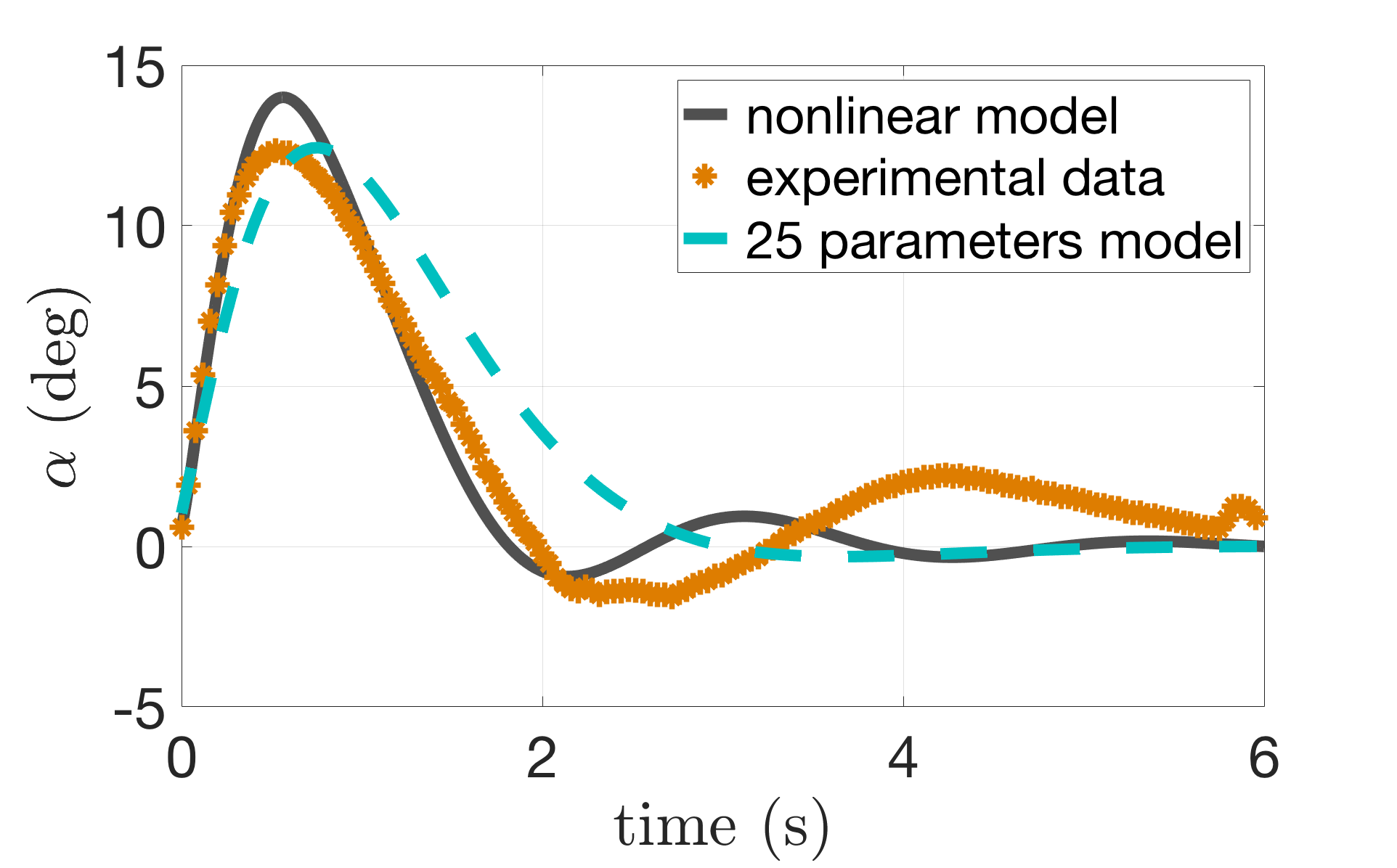}
        \caption{Evolution of the roll angle.}\label{fig:alpha_exper}
    \end{subfigure}
    ~
    \begin{subfigure}[b]{0.45\textwidth}
        \includegraphics[width=\textwidth]{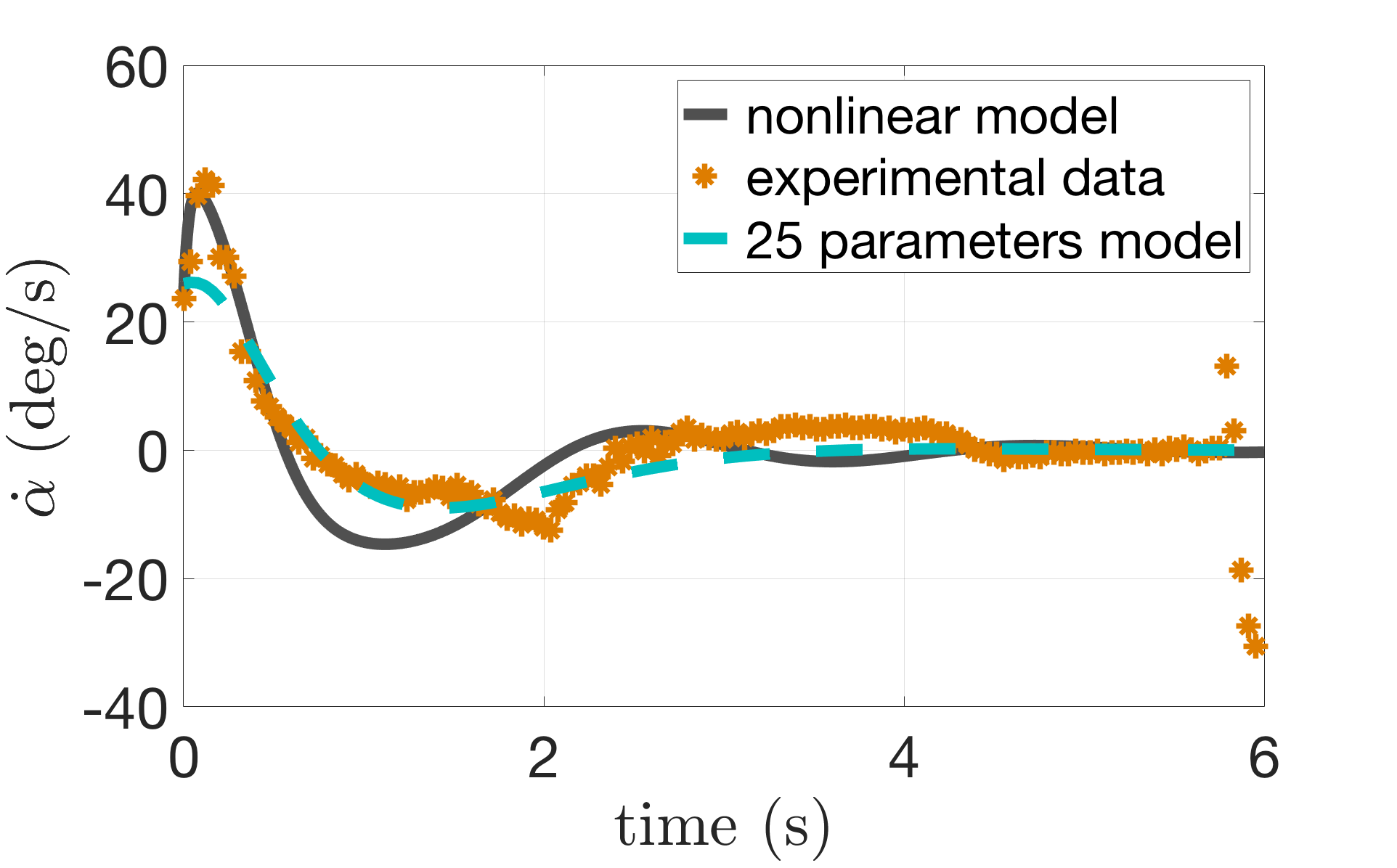}
        \caption{Evolution of the roll angle rate.}\label{fig:dotalpha_exper}
    \end{subfigure}
    
    \begin{subfigure}[b]{0.45\textwidth}
        \includegraphics[width=\textwidth]{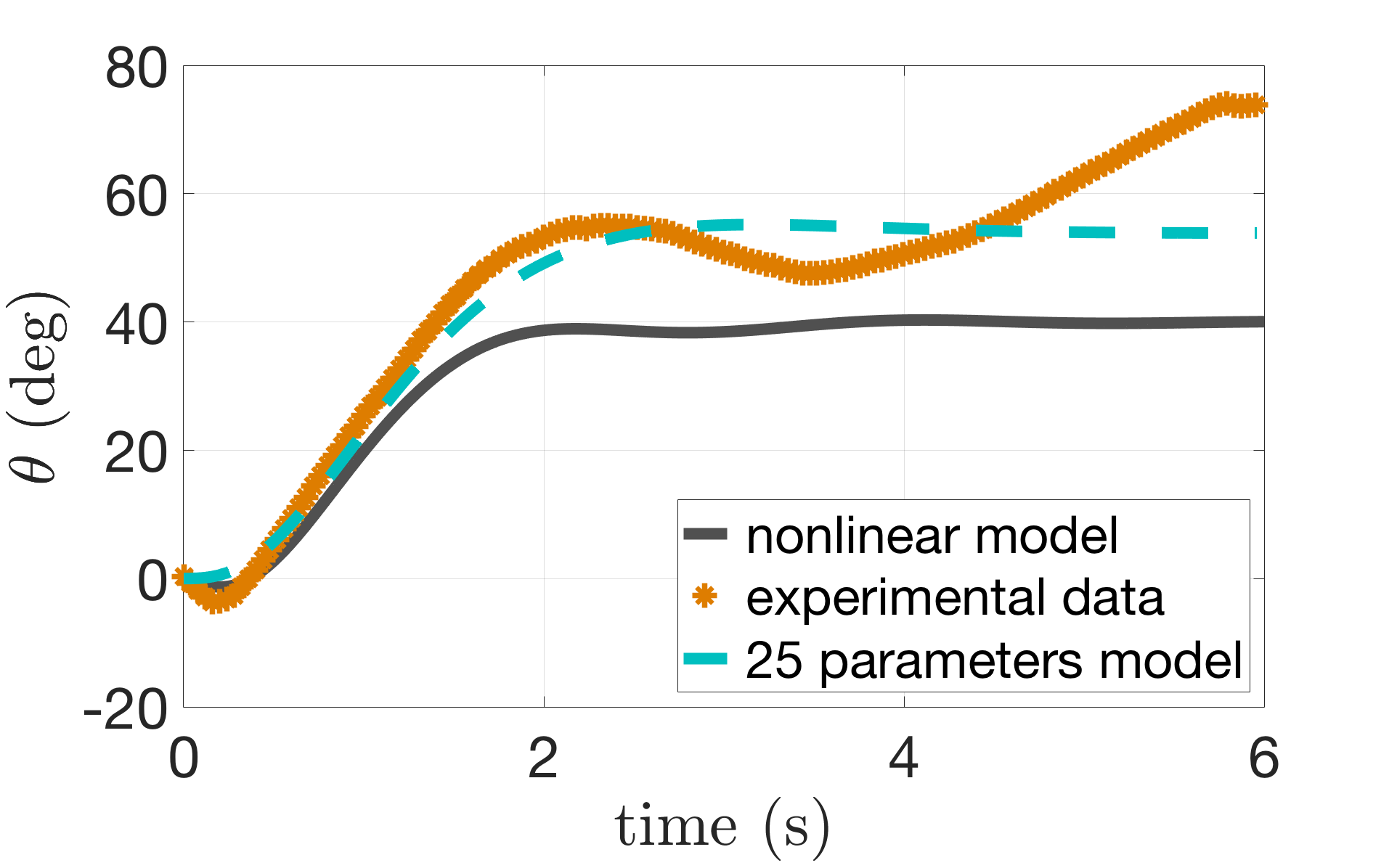}
        \caption{Evolution of the yaw angle.}\label{fig:theta_exper}
    \end{subfigure}
    ~
    \begin{subfigure}[b]{0.45\textwidth}
        \includegraphics[width=\textwidth]{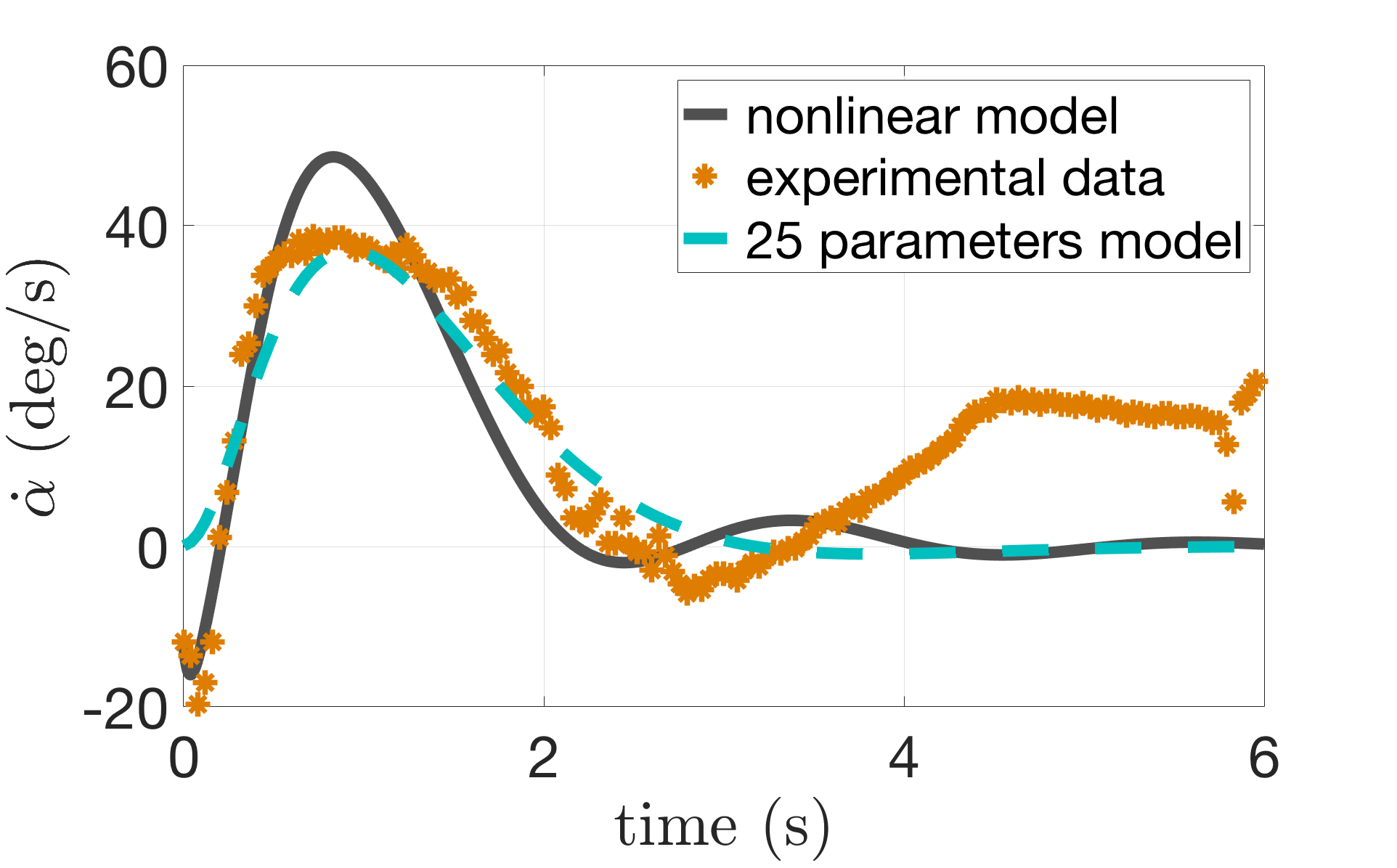}
        \caption{Evolution of the yaw angle rate.}\label{fig:dottheta_exper}
    \end{subfigure}
    \caption{Dynamical behaviour of our nonlinear TMS model (solid line) compared with the experimental data (star dots) measured in~\cite{meijaard:science} and the simulation obtained from the 25 parameters nonlinear model (dashed line) described in~\cite{meijaard:linearized}. Yaw angle and yaw rate are reported with opposite sign from the experimental data according to our definition of the angles.}
    \label{fig:test3}
\end{figure}

In Figure~\ref{fig:test3} we plot the dynamics simulated with our nonlinear TMS model. We compare our results with the data measured during the experiment and the dynamics given by the 25 parameters model defined in~\cite{meijaard:linearized}, which is also reported in the data sheet.

Let us start with the roll angle. From Figure~\ref{fig:alpha_exper} we observe that the measured values are initially fitted by our model. Then, the simulation differs from the measured data, but it still presents the same qualitative behaviour of the measured roll angle. The difference between the measured data and our simulation is mainly due to two factors:
\begin{enumerate}
\item we do not have the precise value of the steering rate as initial condition. Since the system is highly complex, small change in the initial condition change the transient dynamics. We think this is the main reason for the differences from the experimental data shown by the simulation;

\item dissipative forces act on the experimental bicycle and are responsible for the slowing down of the system, but our model does not include friction. According to the report of the experiments in~\cite{kooijman:thesis}, the action of the friction makes the rear speed decrease rapidly, and after $3$ seconds they report a measured rear speed $v_r(3) = \SI{2.4}{\meter\per\second}$.
\end{enumerate}
In the plots we also reported the simulations given by the 25 parameter model. We observe that the simulations given by the linear model give values different from those measured, especially the qualitative behaviour of all the reported angles in significantly different from the one characteristic of the measured data. However, we remark that this simulation is run with initial conditions all zero except for the rear speed and the initial lean rate, which is $\dot\alpha(0) = \SI{0.45}{\radian\per\second}$, as explained in~\cite{meijaard:science}. This may be the reason of such difference.

Regarding the other coordinates, we notice that also the roll angle rate is well predicted by our model, as shown in Figure~\ref{fig:dotalpha_exper}. In particular, our model accurately simulates the jump presented by the measured at the beginning. More differences can be observed in the simulations regarding the yaw angle and its rate, since they do not converge to zero. We also believe that the wrong choice of the initial steering rate influences the values given by the simulation. However, the general picture obtained from our model is representative of experimental data.

\section{Conclusions}

In this paper we studied a nonlinear model for a Two-Mass-Skate bicycle with distributed masses, obtained by combining the simple kinematic structure of the original TMS introduced in~\cite{meijaard:science}, with the geometric effects due to the distributed masses of the rear and front frames.

Using our computer-assisted method for computing the kinetic and potential energies, we have been able to consider these quantities in a symbolic form which allow us to derive explicitly the equations of motion, without resorting to linearise the equations

Since the model obtained is nonlinear, it shows a wide range of dynamical behaviours which strongly depend on the geometric parameters of the model. We observed different asymptotic properties depending on the initial value of the rear speed. In particular, we showed the system can be asymptotically stable, semi-stable, or unstable. We also supposed that different behaviour of the system may be related to the geometric parameters of the model and the existence of circular motions. We proposed few questions suggested by numerical simulations which highlight the main problems we want to address in the future.

Despite there are many works about the bicycle in the literature, we noticed a lack of a realistic and easy nonlinear model which may be used in order to simulate faithfully the real systems dynamics and to design new bicycle. This approach we presented in the paper to overcome this problem can be applied to more general bicycle model, and this is the other direction we want to explore in the future.

\section{Acknowledgements}

We thank Arend Schwab for the email discussion and for helpful suggestions. We gratefully acknowledge fruitful discussions of this work with Enrico Meli.


\end{document}